Scale Invariant Streamline Equations

and

Strings of Singular Vorticity

A.Libin

# 1. Introduction

The importance of coherent structures in a turbulent flow is undoubted. In spite of this, the process of their appearance remains unclear.

In fact, turbulence theory has evolved in the opposite direction during many years. Due to the seminal paper of Kolmogorov in 1941, the emphasis was on statistical concepts of "chaotization" of turbulent flows, while the initial 3-D Navier–Stokes equations remained in the shadows.

The concept that in a developed turbulent flow the Beltrami-type flows are predominant, was initially proposed in the papers by E.Levich and coauthors [6]–[8],[31] and by H.K. Moffat [18],[19],[21]. Though Beltrami flows (anisotropic vector eigenfunctions of the curl operator) are stationary solutions of the Euler equations containing no viscosity terms, modified Beltrami flows with their decay in a viscous fluid, in the absence of external forces, ( solutions of the "force-free Navier–Stokes equations") were found by V.Trkal as early as 1918 [30].

The emergence of domains of helical Beltrami-type flows characterized by the collinearity of the velocity and vorticity vectors is currently firmly established in geophysical observations and numerical simulations of the Navier–Stokes equations [2],[5],[13]–[17],[22],[24],[25]. The fact of formation of large scale Beltrami like helical structures in tornadoes, tropical storms, cloud streets etc. is firmly established by climate observations [26].

In view of this, we here assume the existence of Beltrami-type fluctuations.

To build a mathematical model for the emergence of the coherent structures, we use the idea and technique proposed by G.I. Sivashinskii [29], who regarded large-scale structures as a manifestation of long-wavelength instability of spatially periodic solutions for the Navier–Stokes equations. This approach was used in [10],[11],[27],[28] to study mostly the instability of the linearized 2-D Navier–Stokes equations, sometimes using Gakerkin approximations which could possibly decrease the validity of the obtained analytic result. The 2-D non-linear equations had been investigated for *finite* values of the Reynolds numbers **R**, when the major phenomena is concerned with the trespassing of the so-called critical value **$R_0$** of the Reynolds number [29], whereas CS appear only at large values of **R.**

In [9], the linear stability of the nonstationary Trkal solution [30] for the force-free 2-D Navier–Stokes equations and the stationary Beltrami solution for the forced equations for large values of the Reynolds number $R$ was studied. Although the linearized forced 2-D Navier–Stokes equations with the Beltrami external force are unstable under perturbations with the wavelength $L$ proportional to $R$, it was established that the nonstationary solutions for the linearized force-free Navier–Stokes equations can be unstable under perturbations with an intermediate large wavelength $L$ that is less than $R$; in this case, the order of the quantity $L$ remains unexplained.

In [12] a nonlinear asymptotic analysis of long-wavelength perturbations of the Trkal solution for the force-free 3-D Navier–Stokes equation at large Reynolds numbers $R$ was performed and an asymptotic solution consisting of Beltrami-type flows and terms associated with them was effectively constructed. It turned out that the asymptotic procedure can be implemented in this case only for a single value of the scaling parameter equal to $R^{1/2}$, just as the reduction of the coupling constant in the nonlinear term in renormalized perturbation expansions by $R^{1/2}$, as done by E.Levich. This reduction stems from the assumption that Beltrami fluctuations dominate in a developed turbulent flow [8].

This allows writing equations for plane streamlines that in the quasistationary case turn out to be gradient lines of a function of two variables determined by the initial conditions (actually the energy density); due to emerging upward flow, the resulting 3-D streamlines, as well as the vorticity lines, form invariant under flow 3-D tubes that can be regarded as large-scale structures. H.K.Moffatt envisaged such pictures for the solutions of the Euler equations [20].

As it is explicitly demonstrated here, the crucial point of the whole analysis is the coupling of two large-scale amplitude-modulated anisotropic Beltrami flows with the same eigenvalue of the curl operator ("dual anisotropic Beltrami flows"). Together with the constant orthogonal vector to the pair of dual Beltrami flows linear combinations of the three vectors form solutions of the force-free Navier-Stokes equation, which we call triplets.

No other finite linear combination of linearly independent anisotropic Beltrami flows would yield a solution for the Euler equations. The same is true for the corresponding Trkal solutions for the force-free Navier–Stokes equations. Coupling of two dual "plane" anisotropic Beltrami flows with constant amplitudes yields stationary stratified geometry of streamlines. Variable in space amplitudes yield the emergence of the time depending phase between them, which is unstable at times order $R^{1/2}$ . This time depending phase yields upward velocity and brings the formation of triplets transformed at large Reynolds numbers under long wave-length amplitude perturbation into large scale streamline tubes. At the initial stage at least, inside these tubes the vorticity and the velocity fields are collinear. These streamline tubes are vortex tubes as well. As a result, large scale streamline – vortex tubes are stable at times of order $R^{1/2}$ and vanish at times of order $t \sim R^2$ and thus they might be considered as metastable coherent structures.

The punch of the present endeavour lies in the conviction that the fundamental phenomena of flows in the incompressible liquid in 3-D space are tied to the interplay between explicit time dependence and the inner stream geometry. This is exampled by building of a large scale flow model that stems exclusively from the 3-D Navier-Stokes equations and relates to the accumulated observation data, as well as to the results of computer simulations. Many of them demonstrate the emergence in a flow right from the onset of the regular helical structures, characterised by almost complete alignment of the velocity and of the vorticity vectors, as well as built-in singularity of the vorticity inside streamline-vortex tubes. The proposed model demonstrates that though the flow velocity along the tubes is unstable at times of order $R^{1/2}$, the 3-D tube geometry remains stable almost permanently.

For this purpose, we apply G.I. Sivashinsky's method of multi scaling analysis [29] to the long wavelength perturbations of the so-called Trkal flows at large Reynolds numbers.

## 2. Explicit anisotropic solutions for the force-free Navier-Stokes equations

The Navier-Stokes equations for homogeneous incompressible viscid fluids are:

$$\frac{\partial \vec{u}}{\partial t} + (\vec{u} \cdot \nabla)\vec{u} = -\frac{1}{\rho}\nabla p + \nu \Delta \vec{u} + \vec{f} \qquad (1)$$

where $\vec{u}$ is flow velocity, $p$ is the pressure, $\rho$ is density, assumed to be constant, $\nu$ is the kinematic viscosity, $\vec{f}$ is body force.

The Euler equations for the ideal liquid ($\nu = 0$) are:

$$\frac{\partial \vec{u}}{\partial t} + (\vec{u} \cdot \nabla)\vec{u} = -\frac{1}{\rho}\nabla p + \vec{f} \qquad (2)$$

The so-called Beltrami flow

$$\vec{e}_0(z) = u_0 \begin{pmatrix} \sin(z/d) \\ \cos(z/d) \\ 0 \end{pmatrix}$$

is the solution of the Euler equation (2), while

$$\vec{u}_0 = e^{-\frac{vt}{d^2}} \cdot \vec{e}_0 = u_0 e^{-\frac{vt}{d^2}} \begin{pmatrix} \sin(z/d) \\ \cos(z/d) \\ 0 \end{pmatrix}$$

is the solution of the force-free ($f = 0$) Navier-Stokes equation, where $2\pi d$ is the characteristic space period of the flow, $u_0$ is typical velocity [30].

The Navier-Stokes equation (1) is usually considered in the regularized dimensionless form

$$\begin{cases} \dfrac{\partial \vec{u}}{\partial t} + (\vec{u} \cdot \nabla)\vec{u} = \nabla p + \dfrac{1}{R}\Delta \vec{u} + \dfrac{1}{R}\vec{f}, \\ \operatorname{div} \vec{u} = 0 \end{cases} \qquad (1')$$

where $R = \dfrac{u_0 \cdot d}{v}$ is the so called Reynolds number.

For our purposes we prefer to apply the rot operator to both sides of the last equation.

Due to the well known formula

$$(\vec{u} \cdot \nabla)\vec{u} = [\operatorname{rot} u \times u] + \frac{1}{2}\operatorname{grad}|\vec{u}|^2$$

we get the so called vorticity equation:

$$\frac{\partial(\operatorname{rot}\vec{u})}{\partial t} + \operatorname{rot}[\operatorname{rot}\vec{u} \times \vec{u}] = \frac{1}{R}\Delta(\operatorname{rot}\vec{u}) + \frac{1}{R}\operatorname{rot}\vec{f}$$

Since we intend to investigate the solutions for the force-free Navier-Stokes equation, the last equation is thus reduced to:

$$\begin{cases} \dfrac{\partial (\mathrm{rot}\vec{u})}{\partial t} + \mathrm{rot}[\mathrm{rot}\vec{u} \times \vec{u}] = \dfrac{1}{R}\Delta(\mathrm{rot}\vec{u}) \\ \mathrm{div}\vec{u} = 0 \end{cases} \quad (1")$$

In the dimensionless form the Beltrami flow

$$\vec{e}_1(z) = \begin{pmatrix} \sin z \\ \cos z \\ 0 \end{pmatrix}$$

is a solution for the Euler equation

$$\begin{cases} \dfrac{\partial (\mathrm{rot}\vec{u})}{\partial t} - \mathrm{rot}[\vec{u} \times \mathrm{rot}\vec{u}] = 0, \\ \mathrm{div}\vec{u} = 0 \end{cases} \quad (2')$$

since, obviously

$$\mathrm{rot}\,\vec{e}_1(z) = \vec{e}_1(z),$$

i.e., $\vec{e}_1(z)$ is the eigenvector of the rot operator with the eigenvalue 1.

It is as well obvious that

$$\vec{g}_1(z,t) = A e^{-t/R} \vec{e}_1(z)$$

is a solution for the forceless Navier-Stokes equation (3) (so-called Trkal solution).

In fact,

$$\vec{e}_m(z) = \begin{pmatrix} \sin mz \\ \cos mz \\ 0 \end{pmatrix}, \quad m = 0, \pm 1, \pm 2...$$

is an eigenvector of the rot operator with the eigenvalue $m$, and

$$\vec{g}_m(z,t) = A e^{\frac{-m^2 t}{R}} \vec{e}_m(z),$$

is a solution of the forceless Navier-Stokes equation (1").

On the other hand, it can be easily seen that the same holds for the vectors $\vec{h}_m$, defined as:

$$\vec{h}_m(z) = \begin{pmatrix} \cos mz \\ -\sin mz \\ 0 \end{pmatrix}, \quad m = 0, \pm 1, \pm 2....$$

i.e., $\operatorname{rot} \vec{h}_m(z) = m \vec{h}_m(z)$ and $\vec{v}_m = A e^{\frac{-m^2 t}{R}} \vec{h}_m(z)$ is the solution of the equation (1").

Obviously,

$$[\vec{e}_m(z) \times \vec{e}_n(z)] = \begin{pmatrix} 0 \\ 0 \\ \sin(m-n)z \end{pmatrix}$$

$$[\vec{h}_m(z) \times \vec{h}_n(z)] = \begin{pmatrix} 0 \\ 0 \\ \sin(m-n)z \end{pmatrix} \quad (3)$$

$$[\vec{h}_m(z) \times \vec{e}_n(z)] = \begin{pmatrix} 0 \\ 0 \\ \cos(m-n)z \end{pmatrix}.$$

Clearly, $\vec{e}_m(z)$ and $\vec{h}_m(z)$ are orthogonal. Following P.R. Baldwin and G. M. Townsend [3], we shall call them the dual Beltrami flows.

There are no other linearly independent proper vectors of the rot operator with the proper value *m*, which are anisotropic in *z*.

Consider $2\pi$ periodic three dimensional vector fields with zero divergence, which are anisotropic in *z* (i.e. depending only on *z*).

Then, vectors $\frac{1}{\sqrt{2\pi}}\{\vec{e}_m(z)\}$, $\frac{1}{\sqrt{2\pi}}\{\vec{h}_m(z)\}$ and $\frac{1}{\sqrt{2\pi}}\begin{pmatrix}0\\0\\1\end{pmatrix}$ form the orthogonal basis in a space of square-integrable vector-functions of *z* on $[0, 2\pi]$.

If we seek an anisotropic solution of (1") as a *finite* linear combination of $\{\vec{e}_m(z)\}$ and $\{\vec{h}_m(z)\}$, then, due to (3), the only possible finite linear combinations are:

$$\vec{u} = \gamma_0 \vec{e}_m(z) + \gamma_1 \vec{h}_m(z) + \begin{pmatrix}0\\0\\\delta\end{pmatrix}, \qquad (4)$$

where $\gamma_0, \gamma_1, \delta$ are some functions of time *t*. This follows from the fact that for a given *m*, vectors $\vec{e}_m(z), \vec{h}_m(z)$ and vector $\begin{pmatrix}0\\0\\1\end{pmatrix}$ form a closed set with respect to the cross-product operation:

$$[\vec{e}_m \times \vec{h}_m] = \begin{pmatrix}0\\0\\1\end{pmatrix}; \left[\vec{e}_m(z) \times \begin{pmatrix}0\\0\\1\end{pmatrix}\right] = \vec{h}_m, \left[\vec{h}_m \times \begin{pmatrix}0\\0\\1\end{pmatrix}\right] = -\vec{e}_m.$$

Indeed, if $\vec{u} = \alpha \vec{e}_m + \beta \vec{e}_n$, then in view of the fact that

$$(\vec{g} \cdot \nabla)\vec{f} = \frac{1}{2}\{\text{rot}[\vec{g} \times \vec{f}] + \text{grad}(\vec{g}, \vec{f}) - \vec{f}\,\text{div}\,\vec{g} + \vec{g}\,\text{div}\,\vec{f} - [\vec{f} \times \text{rot}\,\vec{g}] - [\vec{g} \times \text{rot}\,\vec{f}]\},$$

where $(\vec{g}, \vec{f})$ is the scalar product, we have due to (3):

$$(\vec{u} \cdot \nabla)\vec{u} = (n-m) \cdot \alpha\beta [\vec{e}_m \times \vec{e}_n] = (n-m) \cdot \alpha\beta \begin{pmatrix} 0 \\ 0 \\ \sin(n-m)z \end{pmatrix}.$$

Thus, new terms with new space frequencies would appear in equation (1') and $\vec{u}$ cannot be a solution of (1') if $n \neq m$.

We shall call (4) the Beltrami triplet.

Since all the following considerations are valid for any integer $m$, for further deliberations set m=1.

Substitution of (4) into (1") yields

$$\frac{\partial \gamma_1}{\partial t}\vec{h}_1 + \frac{\partial \gamma_0}{\partial t}\vec{e}_1(z) + \delta\gamma_0 \vec{h}_1(z) - \delta\gamma_1 \vec{e}_1(z) = -\frac{\gamma_0}{R}\vec{e}_1(z) - \frac{\gamma_1}{R}\vec{h}_1(z);$$

Hence

$$\begin{cases} \dfrac{\partial \gamma_0}{\partial t} = \delta\gamma_1 - \dfrac{\gamma_0}{R} \\ \dfrac{\partial \gamma_1}{\partial t} = -\delta\gamma_0 - \dfrac{\gamma_1}{R} \end{cases} \quad (5)$$

Multiplying the first equation by $\gamma_0$ and the second one by $\gamma_1$ and summing the result we get

$$\frac{1}{2}\frac{\partial}{\partial t}(\gamma_0^2 + \gamma_1^2) = -\frac{\gamma_0^2 + \gamma_1^2}{R}$$

or

$$\gamma_0^2 + \gamma_1^2 = C_0^2 e^{-\frac{2t}{R}} \; ; \; C_0^2 = \gamma_0^2(0) + \gamma_1^2(0)$$

Thus

$$\begin{aligned} \gamma_0 &= C_0 e^{-\frac{t}{R}} \cos\varphi(t) \\ \gamma_1 &= C_0 e^{-\frac{t}{R}} \sin\varphi(t) \end{aligned} \quad (6)$$

Substitution of (6) into (5) yields

$$\begin{cases} -\sin\varphi \dfrac{\partial\varphi}{\partial t} = \delta\sin\varphi - \dfrac{1}{R}\cos\varphi \\ \cos\varphi \dfrac{\partial\varphi}{\partial t} = -\delta\cos\varphi - \dfrac{1}{R}\sin\varphi \end{cases}$$

Multiplying the first equation by $\sin\varphi$ and the second by $(-\cos\varphi)$ and summing them up, we get

$$-\frac{\partial\varphi}{\partial t} = \delta$$

Hence, the Beltrami triplet can be presented as:

$$u_0 = C_0 e^{-\frac{t}{R}} \cos\varphi(t) \begin{pmatrix} \sin z \\ \cos z \\ 0 \end{pmatrix} + C_0 e^{-\frac{t}{R}} \sin\varphi(t) \begin{pmatrix} \cos z \\ -\sin z \\ 0 \end{pmatrix}$$

$$+ \begin{pmatrix} 0 \\ 0 \\ -\dfrac{\partial\varphi}{\partial t} \end{pmatrix} = \begin{pmatrix} C_0 e^{-\frac{t}{R}} \sin(z+\varphi(t)) \\ -C_0 e^{-\frac{t}{R}} \cos(z+\varphi(t)) \\ -\dfrac{\partial\varphi}{\partial t} \end{pmatrix} \quad (7)$$

Hence, the streamline equations are:

$$\begin{cases} \dot{x} = C_0 e^{-\frac{t}{R}} \sin(z + \varphi(t)) \\ \dot{y} = C_0 e^{-\frac{t}{R}} \cos(z + \varphi(t)) \\ \quad \dfrac{d(z+\varphi)}{dt} = 0 \end{cases} \qquad (6')$$

Thus, if the interaction of two dual plane Beltrami flows yields the time dependent phase $\varphi(t)$, it yields the upward flow as well, depending exclusively on $\varphi(t)$. In other words, the flow becomes three-dimensional only if the coefficients in (4) are not constant.

Vectors $\vec{u}_0$ and $\mathrm{rot}\,\vec{u}_0$ are not collinear and the angle between them is not a small one.

However, $\varphi(t)$ is an arbitrary function of time. This is due to the axial symmetry. To get rid of this indefinite state, we have to break the symmetry.

## 3. The scaling procedure

Let us deviate slightly from the strict anisotropy. Suppose that $C_0, \varphi$ and $\delta$ (and thus $\vec{u}$) in (7) "slowly" depend on $x$ and $y$ and let us extend this scaling to time:

$$\varphi(t) \to \varphi(\varepsilon x, \varepsilon y, \varepsilon t), \quad \frac{1}{R} < \varepsilon < 1$$

where $\varepsilon$ is some small parameter and let us "quench"

$$e^{-\frac{t}{R}} = e^{-\frac{\tau}{\varepsilon R}} = b_0 \approx 1$$

for times under consideration ($t \sim \frac{1}{\varepsilon}$).

Similarly, we shall consider $\gamma_0$ and $\gamma_1$ as functions of $\varepsilon x, \varepsilon y$ and $\varepsilon t$. Hence

$$C_0^2 = \gamma_0^2(0) + \gamma_1^2(0) \to \gamma_0^2(\varepsilon x, \varepsilon y, 0) + \gamma_1^2(\varepsilon x, \varepsilon y, 0) = C_0^2(\varepsilon x, \varepsilon y)$$

We shall assume that all the functions of the new "slow" variables

$$\xi = \varepsilon x \quad (\frac{\partial}{\partial x} \to \varepsilon \frac{\partial}{\partial \xi})$$

$$\eta = \varepsilon y \quad (\frac{\partial}{\partial y} \to \varepsilon \frac{\partial}{\partial \eta})$$

$$\tau = \varepsilon t \quad (\frac{\partial}{\partial t} \to \varepsilon \frac{\partial}{\partial \tau})$$

are periodic in space. Thus, we are dealing with "long-wave perturbations" of a finite amplitude Trkal fluctuation

$$\vec{g}_1(z,t) = A e^{-\frac{t}{R}} \vec{e}_1(z),$$

which is the solution for the force-free Navier-Stokes equations.

We shall seek the perturbed solutions in form (6).

Let us supplement $\gamma_0, \gamma_1$ in the following way:

$$\gamma_0(t) \to A + \gamma_0(\varepsilon x, \varepsilon y, \varepsilon t) = A + \gamma_0(\xi, \eta, \tau)$$
$$\gamma_1(t) \to \gamma_1(\varepsilon x, \varepsilon y, \varepsilon t) = \gamma_1(\xi, \eta, \tau)$$
$$C_0^2(0) \to C_0^2(\xi, \eta) = (\gamma_0(\xi, \eta, 0) + A)^2 + \gamma_1^2(\xi, \eta, 0)$$

Thus, coefficients $\gamma_0(t)$ and $\gamma_1(t)$ in (6) become long wavelength (in $x$ and in $y$) amplitude modulation multipliers for $\vec{e}_1(z)$ and $\vec{h}_1(z)$:

$$\vec{u}_0 \to \vec{u}_0 + \varepsilon \vec{\delta}_1 = (A + \gamma_0(\xi,\eta,\tau))\vec{e}_1(z) + \gamma_1(\xi,\eta,\tau)\vec{h}_1(z) + \varepsilon \begin{pmatrix} 0 \\ 0 \\ -\dfrac{\partial \varphi(\xi,\eta,\tau)}{\partial \tau} \end{pmatrix} \quad (6'')$$

In fact, we already arrived to another form of (7) for the scaled Beltrami triplet:

$$\vec{u}_0 + \varepsilon \vec{\delta}_1 = \begin{pmatrix} C_0(\xi,\eta)\sin(z + \varphi(\xi,\eta,\tau)) \\ C_0(\xi,\eta)\cos(z + \varphi(\xi,\eta,\tau)) \\ 0 \end{pmatrix} + \varepsilon \begin{pmatrix} 0 \\ 0 \\ -\dfrac{\partial \varphi}{\partial \tau} \end{pmatrix} \quad (7')$$

where $C_0^2(\xi,\eta) = (A + \gamma_0(\xi,\eta,0))^2 + \gamma_1^2(\xi,\eta,0)$ is determined by the initial conditions for $\gamma_0$ and $\gamma_1$, i.e. by the initial small long wavelegnth ("noise") modulations of $\vec{e}_1(z)$ and $\vec{h}_1(z)$ amplitudes. Since the angle between $\vec{u}_0$ and rot $\vec{u}_0$ is small and almost proportional to $\varepsilon$, these vectors are close to being collinear.

We have to investigate the rescaled equation (1"):

$$\varepsilon \frac{\partial}{\partial \tau} \operatorname{rot} \vec{u}(\xi,\eta,\tau) + \operatorname{rot}[\operatorname{rot}\vec{u} \times \vec{u}] = \frac{1}{R}\Delta(\operatorname{rot}\vec{u}),$$

where

$$\operatorname{rot}\vec{u}(\xi,\eta,z,\tau) = \operatorname{rot}_z \vec{u} + \varepsilon \operatorname{rot}_{\xi\eta} \vec{u}$$

$$\Delta \vec{u} = \frac{\partial^2 \vec{u}}{\partial z^2} + \varepsilon^2 \left( \frac{\partial^2 \vec{u}}{\partial \xi^2} + \frac{\partial^2 \vec{u}}{\partial \eta^2} \right)$$

i.e. the equation

$$\text{rot}_z[\text{rot}_z u \times u] + \varepsilon\left\{\frac{\partial(\text{rot}_z\vec{u})}{\partial \tau} + \text{rot}_{\xi\eta}[\text{rot}_z\vec{u} \times \vec{u}] + \text{rot}_z[\text{rot}_{\xi\eta}\vec{u} \times \vec{u}] - \frac{1}{R}\frac{\partial^2(\text{rot}_{\xi\eta}\vec{u})}{\partial z^2}\right\} +$$

$$\varepsilon^2\left\{\frac{\partial(\text{rot}_{\xi\eta}\vec{u})}{\partial \tau} - \frac{1}{R}\Delta_{\xi\eta}(\text{rot}_z\vec{u})\right\} - \frac{\varepsilon^3}{R}\Delta_{\xi\eta}(\text{rot}_{\xi\eta}\vec{u}) = \frac{1}{R}\frac{\partial^2(\text{rot}_{\xi\eta}\vec{u})}{\partial z^2} \quad (1''')$$

through the asymptotic expansion of the solution in powers of $\varepsilon$:

$$\vec{u}(\xi,\eta,z,\tau) = \vec{u}_0(\xi,\eta,z,\tau) + \varepsilon\vec{u}_1(\xi,\eta,z,\tau) + \varepsilon^2\vec{u}_2(\xi,\eta,z,\tau) + \ldots \quad .$$

Since the term $\dfrac{1}{R}\dfrac{\partial^2(\text{rot}_z\vec{u})}{\partial z^2}$ in (1''') should be of some order in powers of $\varepsilon$:

$$\varepsilon^k = \frac{1}{R} \text{ for some integer } k.$$

Hence, we consider the asymptotic behavior of $\vec{u}$ with $R$ as a large parameter. In these terms the noncompressibility condition

$$\text{div}\,\vec{u} = 0$$

becomes

$$\text{div}_z\,\vec{u} + \varepsilon\,\text{div}_{\zeta\eta}\vec{u} = 0,$$

i.e.

$$\text{div}_z\,\vec{u}_{k+1} = -\text{div}_{\zeta\eta}\vec{u}_k,$$

Hence

$$\frac{\partial}{\partial z}(\vec{u}_{k+1})_z = -\left(\frac{\partial (\vec{u}_k)_\xi}{\partial \xi} + \frac{\partial (\vec{u}_k)_\eta}{\partial \eta}\right).$$

For *k=0*, we have from (7')

$$\frac{\partial (\vec{u}_1)_z}{\partial z} = -\left(\frac{\partial C_0}{\partial \xi} - C_0 \frac{\partial \varphi}{\partial \eta}\right)\sin(z+\varphi) - \left(\frac{\partial C_0}{\partial \eta} + C_0 \frac{\partial \varphi}{\partial \xi}\right)\cos(z+\varphi),$$

i.e.

$$(\vec{u}_1)_z = \left(\frac{\partial C_0}{\partial \xi} - C_0 \frac{\partial \varphi}{\partial \eta}\right)\cos(z+\varphi) - \left(\frac{\partial C_0}{\partial \eta} + C_0 \frac{\partial \varphi}{\partial \xi}\right)\sin(z+\varphi) + \delta_1(\xi,\eta,\tau)$$

Note, that due to the last formula

$$(\vec{u}_1)_z = (\mathrm{rot}_{\xi\eta}\vec{u}_0(\xi,\eta,z,\tau))_z + \delta_1 \qquad (7'')$$

However, from (7') it follows that

$$\delta_1 = -\frac{\partial \varphi}{\partial \tau}. \qquad (7''')$$

We shall seek other terms of asymptotic expansion in the same form as $\vec{u}_1$ in (7''):

$$\vec{u}_k(\xi,\eta,z,\tau) = \vec{w}_k(\xi,\eta,z,\tau) + \mathrm{rot}_{\xi\eta}\vec{w}_{k-1} + \vec{\delta}_k(\xi,\eta,\tau) \qquad (7'''')$$

where

$\vec{w}_k = \gamma_0^{(k)}(\xi,\eta,\tau)\vec{e}_1(z) + \gamma_1^{(k)}(\xi,\eta,\tau)\vec{h}_1(z)$, i.e,

$$\text{rot}_z \vec{w}_k = \vec{w}_k$$

$$\vec{\delta}_k = \begin{pmatrix} 0 \\ 0 \\ \delta_k(\xi,\eta,\tau) \end{pmatrix}$$

## 4. Scale invariant streamline equations

Now, due to (7') and (7''), we can write the equations for the large scale streamlines:

$$\begin{cases} \dot{\xi} = C_0(\xi,\eta)\sin(z + \varphi(\xi,\eta,\tau)) \\ \dot{\eta} = C_0(\xi,\eta)\cos(z + \varphi(\xi,\eta,\tau)) \end{cases} \quad (8)$$

and

$$\frac{dz}{d\tau} = \left(\frac{\partial C_0}{\partial \xi} - C_0 \frac{\partial \varphi}{\partial \eta}\right)\cos(z+\varphi) - \left(\frac{\partial C_0}{\partial \eta} + C_0 \frac{\partial \varphi}{\partial \xi}\right)\sin(z+\varphi) - \frac{\partial \varphi(\xi,\eta,\tau)}{\partial \tau}$$

However, due to (8),

$$\frac{d\varphi}{d\tau} = \frac{\partial \varphi}{\partial \xi}\cdot\dot{\xi} + \frac{\partial \varphi}{\partial \eta}\cdot\dot{\eta} + \frac{\partial \varphi}{\partial \tau} = C_0 \frac{\partial \varphi}{\partial \xi}\sin(z+\varphi) + C_0 \frac{\partial \varphi}{\partial \eta}\cos(z+\varphi) + \frac{\partial \varphi}{\partial \tau} \quad,$$

Hence

$$\frac{d(z+\varphi)}{d\tau} = \frac{\partial C_0}{\partial \xi}\cos(z+\varphi) - \frac{\partial C_0}{\partial \eta}\sin(z+\varphi). \quad (9)$$

Streamline equations (8) and (9) are asctualy the scaled equations (6'). Equations

(8) are identical to the the first and the second equations in system (6'). We impose requirement of the scaling invariancy of the streamline equations. Hence, we consider the streamlines $(\bar{\xi}(\tau), \bar{\eta}(\tau), \bar{z}(\tau))$, satisfying equations (8) and the equation

$$\frac{d(\bar{z}(\tau) + \varphi(\bar{\xi}(\tau), \bar{\eta}(\tau), \tau))}{d\tau} \equiv 0 \qquad (10)$$

We shall call these streamlines "quasi-stationary trajectories".

Thus, due to (9), we have

$$tg(\bar{z} + \varphi(\bar{\xi}, \bar{\eta}, \tau)) = \frac{\dfrac{\partial C_0}{\partial \xi}}{\dfrac{\partial C_0}{\partial \eta}} \qquad (11)$$

Then

$$\sin(\bar{z} + \varphi(\bar{\xi}, \bar{\eta}, \tau)) = \frac{\dfrac{\partial C_0}{\partial \xi}}{\sqrt{\left(\dfrac{\partial C_0}{\partial \xi}\right)^2 + \left(\dfrac{\partial C_0}{\partial \eta}\right)^2}}$$

$$\cos(\bar{z} + \varphi(\bar{\xi}, \bar{\eta}, \tau)) = \frac{\dfrac{\partial C_0}{\partial \eta}}{\sqrt{\left(\dfrac{\partial C_0}{\partial \xi}\right)^2 + \left(\dfrac{\partial C_0}{\partial \eta}\right)^2}}$$

and we can rewrite (8) for the quasi-stationary trajectories $(\bar{\xi}(\tau), \bar{\eta}(\tau), \bar{z}(\tau))$

$$\begin{cases} \dot{\bar{\xi}} = \dfrac{C_0 \dfrac{\partial C_0}{\partial \xi}}{\sqrt{\left(\dfrac{\partial C_0}{\partial \xi}\right)^2 + \left(\dfrac{\partial C_0}{\partial \eta}\right)^2}} \\ \dot{\bar{\eta}} = \dfrac{C_0 \dfrac{\partial C_0}{\partial \eta}}{\sqrt{\left(\dfrac{\partial C_0}{\partial \xi}\right)^2 + \left(\dfrac{\partial C_0}{\partial \eta}\right)^2}} \end{cases} \quad (12)$$

or, in the vector form

$$\dot{\bar{\Sigma}} = C_0(\bar{\Sigma}) \cdot \frac{\operatorname{grad} C_0(\bar{\Sigma})}{|\operatorname{grad} C_0(\bar{\Sigma})|} \quad (12')$$

where

$$\bar{\Sigma} = \begin{pmatrix} \bar{\xi}(\tau) \\ \bar{\eta}(\tau) \end{pmatrix}$$

Thus, the change of variables ("scaling") and the reqirement of the scaling invariancy enabled the separation of "slow" ($\xi,\eta$) and "fast" (z) variables for the quasi-stationary equations.

Hence, (12') states that the tangent vector to the quasi-stationary trajectory ($\bar{\xi}(\tau),\bar{\eta}(\tau)$) is colinear to the gradient vector of the function $C_0(\xi,\eta)$:

$$\operatorname{grad} C_0(\bar{\xi},\bar{\eta}) = \begin{pmatrix} \dfrac{\partial C_0}{\partial \xi} \\ \dfrac{\partial C_0}{\partial \eta} \end{pmatrix}$$

and the integral lines of the gradient vector field for the function

$$C_0(\xi,\eta) = \left[(Ab_0 + \gamma_0(\xi,\eta))^2 + \gamma_1^2(\xi,\eta)\right]^{1/2}$$

might be considered as "large scale structures", where $\gamma_0(\xi,\eta)$ and $\gamma_1(\xi,\eta)$ are initial small long-wave amplitude modulations ("noise") of the dual Beltrami flows

$$\vec{e}_1(z) = \begin{pmatrix} \sin z \\ \cos z \\ 0 \end{pmatrix} \quad ; \quad \vec{h}_1(z) = \begin{pmatrix} \cos z \\ -\sin z \\ 0 \end{pmatrix}$$

and the projection of the "quasi-stationary" streamline onto the $(\xi,\eta)$ plane is the integral line of the grad $C_0(\xi,\eta)$ vector field.

Consider the evolution of the function

$$C_0(\bar{\xi}(\tau),\bar{\eta}(\tau)) = C_0(\tau)$$

along the quasi-stationary streamline $(\bar{\xi}(\tau),\bar{\eta}(\tau))$:

$$\frac{dC_0(\tau)}{d\tau} = \frac{\partial C_0}{\partial \xi} \cdot \dot{\bar{\xi}}(\tau) + \frac{\partial C_0}{\partial \eta} \dot{\bar{\eta}}(\tau) = \frac{\partial C_0}{\partial \xi} \cdot C_0(\bar{\xi},\bar{\eta}) \sin \bar{w} + \frac{\partial C_0}{\partial \eta} \cdot C_0(\bar{\xi},\bar{\eta}) \cos \bar{w} =$$

$$= C_0(\bar{\xi},\bar{\eta}) \cdot \sqrt{(\frac{\partial C_0}{\partial \xi})^2 + (\frac{\partial C_0}{\partial \eta})^2}$$

where $\bar{w} = \bar{z} + \varphi(\bar{\xi},\bar{\eta},\tau)$.

Or

$$\frac{dC_0(\tau)/d\tau}{C_0(\tau)} = \sqrt{(\frac{\partial C_0}{\partial \xi})^2 + (\frac{\partial C_0}{\partial \eta})^2}$$

Thus,

$$|C_0(\tau)| = c_0 \exp \int_0^\tau \sqrt{(\frac{\partial C_0(\xi(t),\eta(t))}{\partial \xi})^2 + (\frac{\partial C_0(\xi(t),\eta(t))}{\partial \eta})^2} \, dt$$

Since $C_0(\xi,\eta)$ is a bounded function in the whole plane, it follows from the last formula that the quasi-stationary trajectory can only for a finite time remain outside any given neighbourhood of a finite number of the stationary points where the gradient of this function vanishes. Thus, we proved the well known fact that the integral lines of the gradient field connect the "stationary points", where the gradient of $C_0(\xi,\eta)$ vanishes:

$$|\operatorname{grad} C_0(\bar{\xi},\bar{\eta})|=0.$$

## 5. Stability of the quasi stationary solutions

It can be easily seen that $\bar{w}$ is a <u>stable solution</u> of equation (9), with $w = z + \varphi(\xi,\eta,\tau)$.

In fact, set

$$w = \bar{w} + \tilde{w},$$

where $\tilde{w}(\xi,\eta,z,\tau)$ is a small function, and set

$$\xi = \bar{\xi} + \tilde{\xi}, \eta = \bar{\eta} + \tilde{\eta}$$

Then, the linearization of the equations in (8) and (9) yields ($w = z + \varphi$):

$$\begin{cases} \dfrac{d\tilde{\xi}}{d\tau} = \left(\dfrac{\partial C_0}{\partial \xi}\tilde{\xi} + \dfrac{\partial C_0}{\partial \eta}\tilde{\eta}\right)\sin\bar{w} + \tilde{w}C_0(\bar{\xi},\bar{\eta})\cos\bar{w} \\ \dfrac{d\tilde{\eta}}{d\tau} = \left(\dfrac{\partial C_0}{\partial \xi}\tilde{\xi} + \dfrac{\partial C_0}{\partial \eta}\tilde{\eta}\right)\cos\bar{w} + \tilde{w}C_0(\bar{\xi},\bar{\eta})\sin\bar{w} \\ \dfrac{d\tilde{w}}{d\tau} = \dfrac{\partial}{\partial \xi}\left(\dfrac{\partial C_0}{\partial \xi}\tilde{\xi} + \dfrac{\partial C_0}{\partial \eta}\tilde{\eta}\right)\cos\bar{w} - \dfrac{\partial}{\partial \eta}\left(\dfrac{\partial C_0}{\partial \xi}\cdot\tilde{\xi} + \dfrac{\partial C_0}{\partial \eta}\tilde{\eta}\right)\sin\bar{w} - \tilde{w}\left(\dfrac{\partial C_0}{\partial \xi}\sin\bar{w} + \dfrac{\partial C_0}{\partial \eta}\cos\bar{w}\right) \end{cases}$$

But we are interested in the perturbation vector $(\tilde{\xi}(\tau),\tilde{\eta}(\tau),\tilde{w}(\tau))$, orthogonal to the $\operatorname{grad} C_0(\bar{\xi},\bar{\eta})$ vector, i.e, perturbation vectors such that

$$\left(\dfrac{\partial C_0}{\partial \xi}\right)_{\substack{\xi=\bar{\xi}\\ \eta=\bar{\eta}}}\cdot\tilde{\xi} + \left(\dfrac{\partial C_0}{\partial \eta}\right)_{\substack{\xi=\bar{\xi}\\ \eta=\bar{\eta}}}\cdot\tilde{\eta} = 0$$

Consequently,

$$\begin{cases} \dfrac{d\tilde{\xi}}{d\tau} = \tilde{w}C_0(\bar{\xi},\bar{\eta})\cos\bar{w} \\ \dfrac{d\tilde{\eta}}{d\tau} = -\tilde{w}C_0(\bar{\xi},\bar{\eta})\sin\bar{w} \\ \dfrac{d\tilde{w}}{d\tau} = -\left[\dfrac{\partial C_0}{\partial \xi}\sin\bar{w} + \dfrac{\partial C_0}{\partial \eta}\cos\bar{w}\right]\tilde{w} \end{cases}$$

where

$$\operatorname{tg}\bar{w} = \dfrac{\partial C_0}{\partial \xi}\Big/\dfrac{\partial C_0}{\partial \eta}$$

This means that

$$\sin\bar{w} = \dfrac{\partial C_0}{\partial \xi}\cdot\dfrac{1}{\sqrt{(\tfrac{\partial C_0}{\partial \xi})^2 + (\tfrac{\partial C_0}{\partial \eta})^2}}$$

$$\cos\bar{w} = \dfrac{\partial C_0}{\partial \eta}\cdot\dfrac{1}{\sqrt{(\tfrac{\partial C_0}{\partial \xi})^2 + (\tfrac{\partial C_0}{\partial \eta})^2}},$$

and consequently

$$\frac{\partial C_0}{\partial \xi}\sin \overline{w} + \frac{\partial C_0}{\partial \eta}\cos \overline{w} = \sqrt{(\frac{\partial C_0}{\partial \xi})^2 + (\frac{\partial C_0}{\partial \xi})^2}$$

$$\frac{d\tilde{w}}{d\tau} = -\sqrt{(\frac{\partial C_0}{\partial \xi}) + (\frac{\partial C_0}{\partial \eta})^2} \cdot \tilde{w}$$

Since the condition

$$m \leq \sqrt{(\frac{\partial C_0}{\partial \xi})^2 + (\frac{\partial C_0}{\partial \xi})^2} \leq M,$$

is satisfied in the bounded domain (outside arbitrarily small neighborhoods of a finite number of points, at which $\text{grad} C = 0$) we have

$$|\tilde{w}_{(t)}| = C_0 e^{-\int_{\tau_0}^{\tau}\sqrt{(\frac{\partial C_0}{\partial \xi})^2+(\frac{\partial C_0}{\partial \xi})^2}d\tilde{\tau}} \leq C_0 e^{-m(\tau-\tau_0)}$$

Consequently, the streamline projection onto the $(\xi,\eta)$ - plane is a curve asymptotically approaching a "limit curve" defined by the equations for the quasi-stationary streamlines:

$$\dot{\overline{\xi}} = C_0 \cdot \frac{\partial C_0}{\partial \xi} \cdot \frac{1}{|\text{grad} C_0(\overline{\xi},\overline{\eta})|}$$
$$\dot{\overline{\eta}} = C_0 \cdot \frac{\partial C_0}{\partial \eta} \cdot \frac{1}{|\text{grad} C_0(\overline{\xi},\overline{\eta})|}$$

or, in the vector form

$$\dot{\overline{\Sigma}} = C_0(\overline{\Sigma}) \cdot \frac{\text{grad} C_0(\overline{\Sigma})}{|\text{grad} C_0(\overline{\Sigma})|} \quad ,$$

where
$$\overline{\sum} = \begin{pmatrix} \overline{\xi} \\ \overline{\eta} \end{pmatrix},$$

i.e. we got the equations identical with (12) and (12').

Thus, "large scale structures" are formed from these stable $(t \sim \frac{1}{\varepsilon})$ curves in the $(x, y)$ plane. Hence, the question of the streamline behaviour under long wavelength perturbations of the Trkal solution for the forceless Navier-Stokes equation is reduced to the determination of the gradient lines for the function.

$$C_0(\xi,\eta) = \left[ \left( Ab_0 + \gamma_0(\xi,\eta) \right)^2 + \gamma_1^2(\xi,\eta) \right]^{1/2}$$

The stationary points of $C_0(\xi,\eta)$ are either points of maximum ("sources") or minimum values ("syncs") or saddle points.

Every trajectory starts at some maximal point and ends at some minimal point. The saddle point has one incoming and one outgoing trajectory - the separatrices. The plane domain is thus partitioned by the separatrices into invariant sub domains containing the trajectories (plane streamlines), which connect one maximum critical point with one minimum critical point, while there are no other critical points inside these sub domains, i.e. the trajectories inside the sub domains are homotopic.

The assumption of the "long-wavelength perturbation" means that $\gamma_0(\xi, \eta)$ and $\gamma_1(\xi, \eta)$ are two-periodic functions in the $\xi\eta$ plane. Therefore, $\gamma_0(\xi, \eta)$ and $\gamma_1(\xi, \eta)$ are finite trigonometric polynomials in two variables, because all spatial frequencies otherwise participate in any Fourier representation of these functions and the concept of the "long-wavelength perturbation" has no meaning.

## 6. Qualitative (topological) examples

We present two simple examples demonstrating the concept of gradient lines.

The assumption of the "long-wavelength perturbation" means that $\gamma_0(\xi, \eta)$ and $\gamma_1(\xi, \eta)$ are two-periodic functions in the $\xi\eta$ plane. Therefore, $\gamma_0(\xi, \eta)$ and $\gamma_1(\xi, \eta)$ are finite trigonometric polynomials in two variables, because all spatial frequencies otherwise participate in any Fourier representation of these functions and the concept of the "long-wavelength perturbation" has no meaning.

We restrict ourselves to considering the case where only the first harmonics are present in these trigonometric polynomials. Because $\gamma_0(\xi, \eta)$ and $\gamma_1(\xi, \eta)$ as "noisy" initial conditions are small, we can introduce a parameter of smallness $\bar{\varepsilon}$, which, for example, can be the maximum of the absolute values of these functions.

$$C_0 = [(A + \bar{\varepsilon}\bar{\gamma}_0(\xi,\eta))^2 + \bar{\varepsilon}^2 \bar{\gamma}_1^2(\xi,\eta)]^{1/2}$$
$$= [(A^2 + 2\bar{\varepsilon}\bar{\gamma}_0(\xi,\eta))]^{1/2}(1 + \frac{\bar{\varepsilon}^2}{2A}(\bar{\gamma}_0^2 + \bar{\gamma}_1^2) + \cdots)$$

It can be shown using algebraic topology [23] that for sufficiently small $\bar{\varepsilon}$, the picture of the gradient lines for the function $C_0$ is topologically equivalent to that for the function

$$\tilde{C}_0^2 = A^2 + 2\bar{\varepsilon}\bar{\gamma}_0(\xi,\eta)$$

Therefore, the problem of boundary conditions becomes important because the stationary points at which the gradient of $C_0^2$ (and consequently the gradient of $\bar{\gamma}_0$) vanishes depend on the boundary conditions for $\gamma_0(\xi, \eta)$.

As an illustration, we consider the case where the function $\gamma_0(\xi, \eta)$ is $2\pi$-periodic in $\xi$ and $\eta$ and vanish at the boundary of the square $[0, 2\pi] \times [0, 2\pi]$. It is obvious that $\bar{\gamma}_0(\xi,\eta) = K \sin \xi \sin \eta$. This function has five stationary points inside the square (Fig. 1): two maximums $(\pi/2, \pi/2)$ and $(3\pi/2, 3\pi/2)$, two minimums $(\pi/2, 3\pi/2)$ and ($3\pi/2$, $\pi/2$), and one saddle point ($\pi$, $\pi$). The maximum points are sources, and the minimum points are sinks. Two separatrices pass through the saddle point (Fig.1).

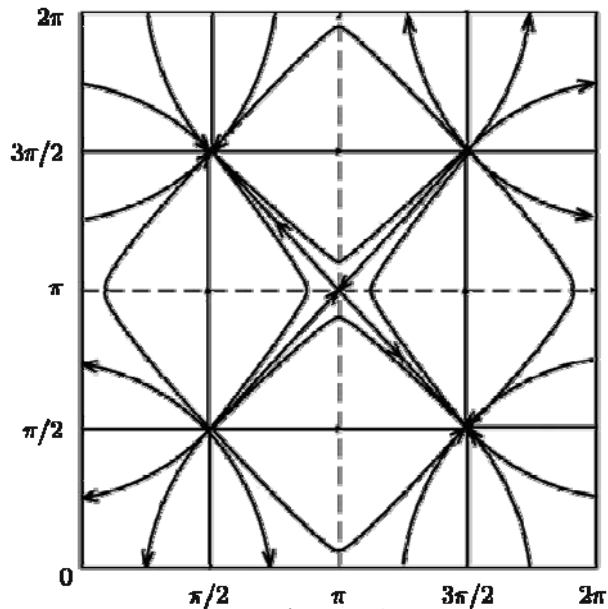
Figure 1

The example of $2\pi$-periodic boundary conditions on the boundaries of the square [0, $2\pi$] × [0, $2\pi$] for the function $\gamma_0(\xi, \eta)$ (i.e., $\gamma_0 (\xi, \eta)$ is considered on the torus) is less obvious.

V. Arnold [1] wrote these functions in the form

$$\bar{\gamma}_0(\xi,\eta) = a \cos \xi + b \sin \xi + c \cos \eta + d \sin \eta + p \cos(\xi+\eta) + q \sin(\xi+\eta)$$

and proved that they have six stationary points and they allow two different topological pictures (with respect to the diffeomorphism group of the torus) for the level lines.

These two pictures (and consequently the gradient line pictures) are determined by the structure of the six stationary points: :
- two maximal points one maximal points, three saddle points and two minimal points;
- three saddle points and one minimal point.

The corresponding picture of the gradient lines for first case is presented in Figure 2.

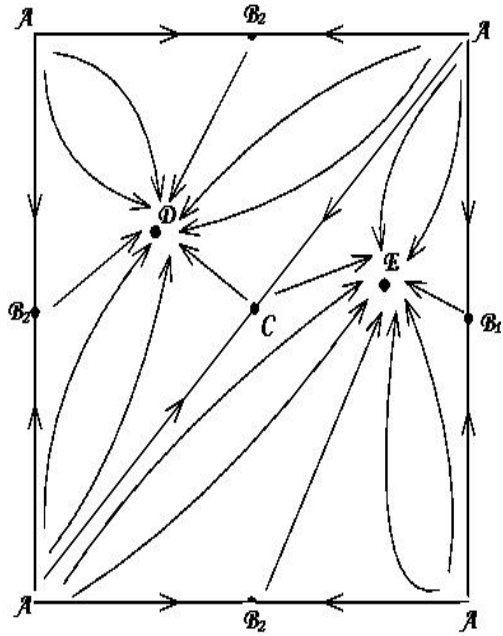

Figure 2

It can be easily seen that the plane will be divided into "curved polygons", the stationary points being the polygon's vertices, while the polygon's sides are the separatrices, the gradient lines that separate subsets of homotopic gradient lines. Each polygon is an invariant set under the gradient flow.

**7. Explicit time dependence and the emergence of the upward flow**

In order to elucidate the three dimensional behaviour of the quasi-stationary trajectories, we have to elaborate other terms of the asymptotic expansion

$$\vec{u}(\xi,\eta,z,\tau) = \vec{u}_0(\xi,\eta,z,\tau) + \varepsilon\vec{u}_1 + \varepsilon^2\vec{u}_2 + ...$$
$$\text{div}\,\vec{u} = 0 \quad (13)$$

We shall seek other terms in the same form as $\vec{u}_1$ in (7"):

$$\vec{u}_k(\xi,\eta,z,\tau) = \vec{w}_k(\xi,\eta,z,\tau) + \text{rot}_{\xi\eta}\vec{w}_{k-1} + \vec{\delta}_k(\xi,\eta,\tau) \quad (14)$$

where

$$\vec{w}_k = \gamma_0^{(k)}(\xi,\eta,\tau)\vec{e}_1(z) + \gamma_1^{(k)}(\xi,\eta,\tau)\vec{h}_1(z), \text{ i.e,}$$

$$\text{rot}_z \vec{w}_k = \vec{w}_k$$

$$\vec{\delta}_k = \begin{pmatrix} 0 \\ 0 \\ \delta_k(\xi,\eta,\tau) \end{pmatrix}$$

For the quasi-stationary trajectory (10) we have:

$$\overline{z}(\tau) = C_0(\xi_0,\eta_0,\tau_0) - \varphi(\overline{\xi}(\tau),\overline{\eta}(\tau),\tau), \; \xi_0 = \overline{\xi}(\tau_0), \eta_0 = \overline{\eta}(\tau_0) \qquad (14')$$

i.e., one has to find $\varphi(\xi,\eta,\tau)$.

To do this let us integrate equation (1'''')

$$\varepsilon \frac{\partial}{\partial \tau}\text{rot}\,\vec{u}(\xi,\eta,\tau) + \text{rot}[\text{rot}\,\vec{u} \times \vec{u}] = \frac{1}{R}\Delta(\text{rot}\,\vec{u}),$$

over the period in $z$, substituting $\vec{u}$ from (13) and $\vec{u}_k$ from (14) and taking into account, that

$$\text{rot}\,\vec{u} = \text{rot}_z \vec{u} + \varepsilon \text{rot}_{\xi\eta} \vec{u}$$

Hence,

$$\varepsilon^2 \text{rot}_{\xi\eta}\left(\frac{\partial}{\partial \tau}\int_0^{2\pi} \vec{u}(\xi,\eta,z,\tau)dz\right) + \varepsilon \text{rot}_{\xi\eta}\left(\int_0^{2\pi}[\text{rot}_z \vec{u} \times \vec{u}]dz\right) +$$

$$+ \varepsilon^2 \text{rot}_{\xi\eta}\left(\int_0^{2\pi}[\text{rot}_{\xi\eta} \vec{u} \times \vec{u}]dz\right) = \frac{\varepsilon^3}{R}\text{rot}_{\xi\eta}\int_0^{2\pi}(\text{rot}_{\xi\eta}\vec{u})dz$$

In [9] it was demonstrated that for the **linearised problem** of the long wavelength stability of the Trkal flows there exists such (indefinite) large length scale

L=$1/\varepsilon$, which lies between 1 and R, that the perturbations of the linearised scaled problem might be unstable. Consequently, the thrust of the present investigation lies in the assumption

$$\varepsilon > \frac{1}{R}.$$

Hence, the term in the right-hand side is of order $\varepsilon^4$ or of higher orders in powers of $\varepsilon$. We can drop $\text{rot}_{\xi\eta}$ in both sides of the last equation. For $\varepsilon^0, \varepsilon^1$ and $\varepsilon^2$ approximations we get:

$$\varepsilon^0 : \vec{u}_0 = C_0 \begin{pmatrix} \sin(z+\varphi) \\ \cos(z+\varphi) \\ 0 \end{pmatrix} \to [\text{rot}_z \vec{u}_0 \times \vec{u}_0] \equiv 0;$$

$$\varepsilon^1 : \int_0^{2\pi} \left\{ [\vec{w}_1 \times \vec{u}_0] + [\vec{u}_0 \times (\vec{w}_1 + \text{rot}_{\xi\eta}\vec{u}_0 + \vec{\delta}_1)] \right\} dz =$$

$$+ \int_0^{2\pi} [\text{rot}_{\xi\eta}\vec{u}_0 \times \vec{u}_0] dz = 0;$$

$$\varepsilon^2 : \int_0^{2\pi} \left( \frac{\partial}{\partial \tau} \vec{u}_1 \right) dz + \int_0^{2\pi} \left\{ [\text{rot}_{\xi\eta}\vec{u}_2 \times \vec{u}_0] + [\text{rot}_z \vec{u}_1 \times \vec{u}_1] + [\text{rot}_z \vec{u}_0 \times \vec{u}_2] \right\} dz +$$

$$+ \int_0^{2\pi} [\text{rot}_{\xi\eta}\vec{u}_1 \times \vec{u}_0] dz + \int_0^{2\pi} [\text{rot}_{\xi\eta}\vec{u}_0 \times \vec{u}_1] dz = 0$$

$$\int_0^{2\pi} \left( \frac{\partial}{\partial \tau} \vec{u}_1 \right) dz + \int_0^{2\pi} \left\{ [\vec{w}_2 \times \vec{u}_0] + [\vec{w}_1 \times (\vec{w}_1 + \text{rot}_{\xi\eta}\vec{u}_0 + \vec{\delta}_1)] + [\vec{u}_0 \times \vec{w}_2] \right\} dz +$$

$$+ \int_0^{2\pi} \left[ (\text{rot}_{\xi\eta}\vec{w}_1 + \text{rot}_{\xi\eta}(\text{rot}_{\xi\eta}\vec{u}_0) + \text{rot}_{\xi\eta}\vec{\delta}_2) \times \vec{u}_0 \right] dz + \int_0^{2\pi} [\text{rot}_{\xi\eta}\vec{u}_0 \times (\vec{w}_1 + \text{rot}_{\xi\eta}\vec{u}_0 + \vec{\delta}_1)] dz = 0$$

From the anti-commutating properties of the cross product and the fact that $\delta_1(\xi,\eta,\tau)$ and $\delta_2(\xi,\eta,\tau)$ are not depending on $z$, and thus all the integrals with $\vec{\delta}_1$ and $\vec{\delta}_2$ are vanishing, we get

$$\int_0^{2\pi} \frac{\partial \vec{u}_1}{\partial \tau} dz + \int_0^{2\pi} [(\text{rot}_{\xi\eta}\text{rot}_{\xi\eta}\vec{u}_0) \times \vec{u}_0] dz = 0$$

We are seeking $\vec{u}_1$ in the form (14): $\vec{u}_1 = \vec{w}_1(z) + \mathrm{rot}_{\xi\eta}\vec{u}_0 + \vec{\delta}_1$.

Integrals over $z$ of the first two terms vanish.

Direct evaluation of $\mathrm{rot}_{\xi\eta}\mathrm{rot}_{\xi\eta}\vec{u}_0(z)$ and further integration in $z$ yield:

$$\frac{\partial \delta_1}{\partial \tau} = \frac{1}{2}C_0^2(\xi,\eta)\Delta_{\xi\eta}\varphi(\xi,\eta,\tau) + \left(\frac{\partial C_0}{\partial \xi}\frac{\partial \varphi}{\partial \xi} + \frac{\partial C_0}{\partial \eta}\frac{\partial \varphi}{\partial \eta}\right)\cdot C_0$$

Hence, due to (7''')

$$\frac{\partial^2 \varphi}{\partial \tau^2} = -\frac{C_0^2(\xi,\eta)}{2}\Delta_{\xi\eta}\varphi - C_0\left(\frac{\partial C_0}{\partial \xi}\frac{\partial \varphi}{\partial \xi} + \frac{\partial C_0}{\partial \eta}\frac{\partial \varphi}{\partial \eta}\right) \qquad (15)$$

with the initial conditions

$$\varphi(\xi,\eta,0) = \mathrm{arctg}\frac{\gamma_1(\xi,\eta,0)}{\gamma_0(\xi,\eta,0) + A}$$

$$\frac{\partial}{\partial \tau}\varphi(x,y,0) = \delta_0(\xi,\eta)$$

Thus, we have a Cauchy problem (with periodic boundary conditions) for an elliptic partial differential equation - the classic example of an "ill posed problem", i.e., the case of instability in time of the phase $\varphi(\xi,\eta,\tau)$, as well as of the upward velocity. Since (15) had been elaborated without involvement of the viscous terms, we are dealing with the *Eulerian phase instability*. The negative sign before the Laplacian in (15) is usually considered as the manifestation of the so-called "negative viscosity" [26-28] at times $\tau \sim 1, (t \sim \frac{1}{\varepsilon})$. In case the initial upward flow is absent, $\varphi(\xi,\eta,\tau)$ is non trivial only if the initial long wavelength amplitude modulation $\gamma_1(\xi,\eta,0) = \gamma_*(\xi,\eta)$ of the dual flow is not vanishing, i.e. both explicit time dependence of the flow as well as the emergence of the upward flow, which makes the flow essentially three dimensional, are due to the initial space gradient of the second coefficient in (7').

As a result, we have found *z(t)* in (14').

In fact, we may speak about the "pseudo chaotic" behaviour of $z(t)$, due to (15),as against stable structures, defined by (12'). The same function $C_0(\xi,\eta)$ determines (12') as well as (14'), i.e. order and "pseudo chaos" emerge from the same cause: small ("noise") long wavelength amplitude modulation in $(\xi,\eta)$–plane of a pair of dual anisotropic Beltrami flows [5]. This is due to the fact that in the initial conditions in (15) enter into the function $C_0(\xi,\eta)$, which appears as the coefficient before the highest order derivative and thus a small variation of in initial conditions can cause large variation of the solution (in fact, this is the so called "Hadamard example" of instability with respect to initial conditions of the Cauchy problem for the elliptic equations) as well as variation of $C_0(\xi,\eta)$. Thus, variation of the initial conditions in (15) can cause instability of the solution and of vertical velocity, while the solutions of (12) will remain stable.

It can be easily seen from (13) and (14) that

$$\vec{u} - \operatorname{rot}\vec{u} = \varepsilon \vec{\delta}_1 + 0(\varepsilon^2),$$

where

$$\operatorname{rot} \to \varepsilon \operatorname{rot}_{\xi\eta} + \operatorname{rot}_z,$$

i.e. *velocity and vorticity are almost collinear,* at least, at initial stage.

The last term in the right hand side contains all the terms of the higher orders in the asymptotic expansion, while the first term in the right hand-side is a vector, which is parallel to the z –axis. Thus, up to terms of order $\varepsilon^2$, both velocity and vorticity vectors belong to a tangent plane of a vertical surface, which contains a curve in the (*x, y*) plane, determined by the equations (12). This is true even when the first term in the right-hand side of the last equation is not small, i.e. when velocity and vorticity cease to be almost collinear. Thus, this surface might be considered as a "streamline sheet" as well as a "vortex sheet". Streamline sheets, which contain homotopic quasi stationary trajectories of the same subdomain in (*x, y*) plane, connecting two fixed stationary points, form the invariant three dimensional domains, called "streamline tubes". Clearly, up to terms of order $\varepsilon^2$, "streamline tubes" are at the same time "vortex tubes". These streamline–vortex tubes will be stable at times $\tau \sim 1, (t \sim \frac{1}{\varepsilon})$.

This unveils the 3-D picture of a streamline-vortex tube, which is in fact a "curved upright prism" with vertical edges, growing from the stationary points, based on the plane "curved polygon" made by the separatrices

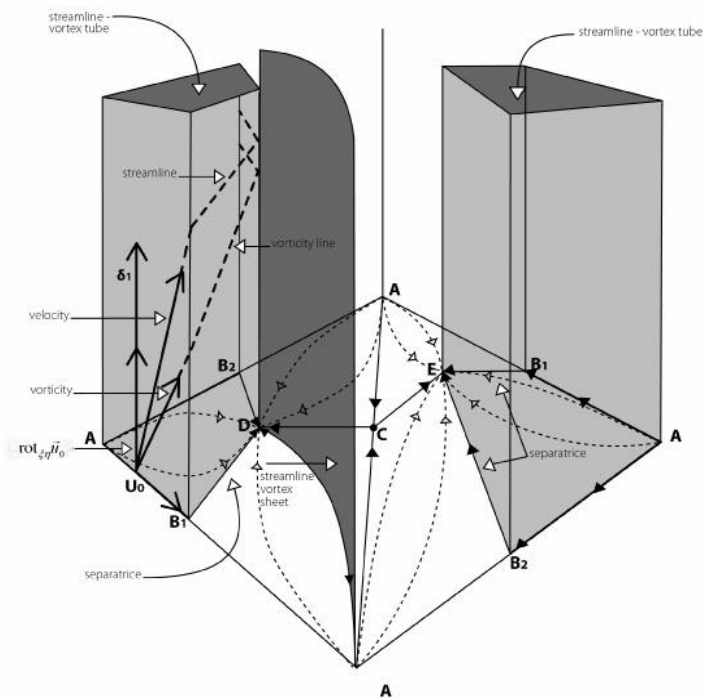

Figure 3

Streamline-vortex tubes.

Figure 3 shows the 3-D picture emerging from the plane picture on Figure 2.

In fact, since $\varphi(\xi,\eta,\tau)$ can be determined as a solution of the Cauchy problem (15), we proved that all the terms in the asymptotic expansion (13)-(14) can be consequently found.

## 8. Determination of the asymptotic expansion parameter

The gradient lines equations (12') relate to the quasi-stationary streamlines for the vector field $\vec{u}_0$ in the $(\xi,\eta)$-plane. As it was already mentioned, it can be proved by means of algebraic geometry that small perturbation of the gradient stream does not change the topology of the gradient lines in the $(\xi,\eta)$ - plane [23].

We have established the equations (12) from equations for $\varepsilon^0$ approximation in the $(\xi,\eta)$ - plane. In order to prove that other terms in asymptotic expansion would not change the topological picture of the gradient flow, we have to prove that $|\vec{w}_1|$ is bounded.

In order to elaborate $\vec{w}_1$ (the first term in $\vec{u}_1$) we have to determine for which integer $\kappa$

$$\varepsilon^\kappa = R^{-1},$$

Suppose first, $\kappa > 2$.

This yields for the $\varepsilon^2$ approximation

$$\frac{\partial \vec{w}_1}{\partial \tau} + [\vec{w}_1 \times \vec{\delta}_1] + [\vec{u}_0 \times \vec{\delta}_2] = 0 \qquad (16)$$

To elucidate $\vec{w}_1$, we get the equations

$$\begin{cases} \dfrac{\partial \gamma_0^{(1)}}{\partial \tau} = \delta_1 \gamma_1^1 + \delta_2 \gamma_1^0 \\ \dfrac{\partial \gamma_1^{(1)}}{\partial \tau} = -\delta_1 \gamma_0^1 - \delta_2(\gamma_0^0 - A) \end{cases} \quad (17)$$

Consider the auxiliary system

$$\begin{cases} \dfrac{\partial \tilde{\gamma}_0^{(1)}}{\partial \tau} = \delta_1 \tilde{\gamma}_1^1 \\ \dfrac{\partial \tilde{\gamma}_1^{(1)}}{\partial \tau} = -\delta_1 \tilde{\gamma}_0^1 \end{cases} \quad (17')$$

Obviously

$$(\tilde{\gamma}_0^{(1)})^2 + (\tilde{\gamma}_1^{(1)})^2 = \tilde{C}_0^2(\xi, \eta),$$

Thus

$$\tilde{\gamma}_0^{(1)} = \tilde{C}_0 \cos \tilde{\varphi}(\xi, \eta, \tau)$$
$$\tilde{\gamma}_1^{(1)} = \tilde{C}_0 \sin \tilde{\varphi}(\xi, \eta, \tau).$$

Substituting this into (17'), we get

$$\delta_1 = -\dfrac{\partial \tilde{\varphi}}{\partial \tau}$$

Thus, in view of (7")

$$\tilde{\varphi} = \varphi(\xi, \eta, \tau) + \tilde{\varphi}_0(\xi, \eta)$$

Hence, we shall seek the solution of (17) as

$$\gamma_0^{(1)}(\xi,\eta,\tau) = \tilde{\tilde{C}}(\xi,\eta,\tau)\cos\tilde{\varphi}(\xi,\eta,\tau)$$

$$\gamma_1^{(1)}(\xi,\eta,\tau) = \tilde{\tilde{C}}(\xi,\eta,\tau)\sin\tilde{\varphi}(\xi,\eta,\tau)$$

Substituting this into (17), we get

$$\cos\tilde{\varphi}\frac{\partial \tilde{\tilde{C}}}{\partial \tau} = \delta_2 \gamma_1^0$$

$$\sin\tilde{\varphi}\frac{\partial \tilde{\tilde{C}}}{\partial \tau} = -\delta_2 (A+\gamma_0^0)$$

Multiplying the first equation by $A+\gamma_0^0 = C_0\cos\varphi$ and the second one by $\gamma_1^0 = C_0\sin\varphi$, we get

$$C_0 \cos(\tilde{\varphi}-\varphi)\frac{\partial \tilde{\tilde{C}}}{\partial \tau} = 0$$

or

$$C_0 \cos\varphi_0 \cdot \frac{\partial \tilde{\tilde{C}}}{\partial \tau} = 0$$

Then,

$$\delta_2 = 0$$

Hence

$$\tilde{C}^2(\xi,\eta) = (\gamma_1^{(1)}(\xi,\eta,0))^2 + (\gamma_0^{(1)}(\xi,\eta,0))^2$$

But since the amplitudes in the initial conditions are small ("noise"), they enter into initial conditions for the zero approximation, the initial conditions for higher order terms in asymptotic expansion (13) are zero.

Thus

$$\tilde{C}_0(\xi,\eta) \equiv 0$$

and all the terms in the asymptotic expansion in form (14) for $k > 1$ are vanishing. However, $\vec{u}_0 + \varepsilon \vec{u}_1$ is not an exact solution of (13).

Therefore, if we suppose that $\kappa > 2$, the solution of (13) in form (14) does not exist.

Thus $k = 2$ and

$$\varepsilon = R^{-1/2}.$$

Now we have to prove that the first term of asymptotic expansion (13) in form (14) does not change the plane topological picture determined by (12) and depicted in Fig. 2 and Fig. 3.

The plane part of the first term in (13) is

$$\frac{1}{\sqrt{R}} \vec{w}_1.$$

Therefore, we have to prove that $|\vec{w}_1|$ is bounded.

For $\kappa = 2$, the term $\frac{1}{R}\Delta_z(\text{rot}_z \vec{u}_0)$ appears in the equations for $\varepsilon^2$ approximation and we get the following equation for $\vec{w}_1$:

$$\frac{\partial \vec{w}_1}{\partial \tau} + [\vec{u}_0 \times \vec{\delta}_2] + [\vec{w}_1 \times \vec{\delta}_1] + \vec{u}_0 = 0$$

To elucidate $\vec{w}_1$, we get the equations

$$\begin{cases} \dfrac{\partial \gamma_0^{(1)}}{\partial \tau} = \delta_1 \gamma_1^1 + \delta_2 \gamma_1^0 - \gamma_0^0 \\ \dfrac{\partial \gamma_1^{(1)}}{\partial \tau} = -\delta_1 \gamma_0^1 - \delta_2 (A + \gamma_0^0) - \gamma_1^0 \end{cases} \qquad (17'')$$

Consider the auxiliary system

$$\begin{cases} \dfrac{\partial \tilde{\gamma}_0^{(1)}}{\partial \tau} = \delta_1 \tilde{\gamma}_1^1 \\ \dfrac{\partial \tilde{\gamma}_1^{(1)}}{\partial \tau} = -\delta_1 \tilde{\gamma}_0^1 \end{cases} \qquad (17''')$$

Obviously

$$(\tilde{\gamma}_0^{(1)})^2 + (\tilde{\gamma}_1^{(1)})^2 = \tilde{C}_0^2(\xi, \eta),$$

Thus

$$\tilde{\gamma}_0^{(1)} = \tilde{C}_0 \cos \tilde{\varphi}(\xi, \eta, \tau)$$
$$\tilde{\gamma}_1^{(1)} = \tilde{C}_0 \sin \tilde{\varphi}(\xi, \eta, \tau),$$

Substituting this into (17''), we get

$$\delta_1 = -\dfrac{\partial \tilde{\varphi}}{\partial \tau}$$

Thus, in view of (7'')

$$\tilde{\varphi} = \varphi(\xi, \eta, \tau) + \tilde{\varphi}_0(\xi, \eta)$$

Hence, we shall seek the solution of (17'') as

$$\gamma_0^{(1)}(\xi, \eta, \tau) = \tilde{\tilde{C}}(\xi, \eta, \tau) \cos \tilde{\varphi}(\xi, \eta, \tau)$$
$$\gamma_1^{(1)}(\xi, \eta, \tau) = \tilde{\tilde{C}}(\xi, \eta, \tau) \sin \tilde{\varphi}(\xi, \eta, \tau)$$

Substituting this into (17''), we get

$$\cos\tilde{\varphi}\frac{\partial \tilde{\tilde{C}}}{\partial \tau} = \delta_2 \gamma_1^0 - \gamma_0^0$$

$$\sin\tilde{\varphi}\frac{\partial \tilde{\tilde{C}}}{\partial \tau} = -\gamma_1^0 - \delta_2(A+\gamma_0^0)$$

Multiplying the first equation by $A + \gamma_0^0 = C_0 \cos\varphi$ and the second one by $\gamma_1^0 = C_0 \sin\varphi$, we get

$$C_0 \cos(\tilde{\varphi} - \varphi)\frac{\partial \tilde{\tilde{C}}}{\partial \tau} = -[(\gamma_1^0)^2 + (A+\gamma_0^0)^2] + A(A+\gamma_0^0)$$

or

$$C_0 \cos\varphi_0 \cdot \frac{\partial \tilde{\tilde{C}}}{\partial \tau} = -C_0^2 + AC_0 \cos\varphi$$

Hence

$$\frac{\partial \tilde{\tilde{C}}}{\partial \tau} = -\frac{C_0}{\cos\varphi_0} + \frac{A\cos\varphi}{\cos\varphi_0},$$

and

$$\tilde{\tilde{C}}(\xi,\eta,\tau) = \frac{1}{\cos\varphi_0}\int_0^\tau [A\cos\varphi(\xi,\eta,s) - C_0(\xi,\eta)]ds + B_0(\xi,\eta).$$

However, as in the case of $k > 2$,

$$\tilde{\tilde{C}}^2(\xi,\eta,0) = (\gamma_0^1(\xi,\eta,0))^2 + (\gamma_1^1(\xi,\eta,0))^2 \equiv 0,$$

since the initial conditions for the higher order terms in the asymptotic expansion in powers of a small parameter for the **small "noise"** are zero.

Therefore

$$B_0(\xi,\eta) \equiv 0$$

Hence,

$$|\vec{w}_1| = |\tilde{\tilde{C}}(\xi,\eta,\tau)| = \int_0^\tau \left|\frac{A\cos\varphi(\xi,\eta,s) - C_0(\xi,\eta)}{\cos\varphi_0(\xi,\eta)}\right| ds \le$$

$$\le \frac{\max_{(\xi,\eta)} C_0(\xi,\eta) + A}{|\cos\tilde{\varphi}_0|(1-\varepsilon_0)} \cdot \tau = M\tau.$$

Thus, for $\tau \sim 1$ ($t \sim \sqrt{R}$), $|w_1|$ is bounded.

The same can be proved for the plane components of higher order terms in (13).

Thus, we completed the asymptotic procedure for $\varepsilon^0$ and $\varepsilon^1$. Obviously, it can be done successively for any $k$, while corresponding equations for $k > 1$ will include as free terms $\gamma_n^{(0)}, \gamma_n^{(1)}, \delta_{n+1}$ for $k > n$, since these functions had been elucidated on the earlier stages and are considered as known.

Hence, $R^{1/2}$ is the characteristic size of the invariant domains in the (*x,y*) plane or the characteristic "diameter" of the three dimensional invariant streamline-vortex tubes. For times $t \sim R^{1/2}$ the phase $\varphi(\xi,\eta,\tau)$ between two dual Beltrami flows satisfies equation (15) with negative Laplacian in the right–hand side, i.e. the large scale viscosity becomes negative, while the upward velocity becomes unstable.

## 9. Decay of the upward flow

As it was mentioned several times, equation (15) clearly indicated on the growth of the upward velocity for times $t \sim R^{1/2}$. To elucidate the behaviour of the upward velocity at very large times $t \gg R$, we should consider the new change of variables, which is identical with the first scaling in space variables, but it is different in time scaling:

$$\xi = \frac{x}{\sqrt{R}}$$

$$\eta = \frac{y}{\sqrt{R}}$$

$$\tau_1 = \frac{t}{R^2}$$

Substitution into (7) yields for $t \gg R$

$$u_0 = C_0 e^{-\frac{t}{R}} \cos\varphi(t) \begin{pmatrix} \sin z \\ \cos z \\ 0 \end{pmatrix} + C_0 e^{-\frac{t}{R}} \sin\varphi(t) \begin{pmatrix} \cos z \\ -\sin z \\ 0 \end{pmatrix}$$

$$+ \begin{pmatrix} 0 \\ 0 \\ -\frac{\partial \varphi}{\partial t} \end{pmatrix} = \begin{pmatrix} C_0 e^{-\frac{t}{R}} \sin(z + \varphi(t)) \\ -C_0 e^{-\frac{t}{R}} \cos(z + \varphi(t)) \\ -\frac{\partial \varphi}{\partial t} \end{pmatrix} \to \begin{pmatrix} 0 \\ 0 \\ \delta(\xi, \eta, \tau_1) \end{pmatrix}$$

Thus for large $R$, the scaled solution is a vector field that is parallel to the $z$-axis and does not depend on $z$. On the other hand, the newly rescaled equation (1") is:

$$\frac{1}{R^{5/2}} \frac{\partial}{\partial \tau_1}(\text{rot}_{\xi\eta}\vec{u}_0) + \text{rot}_z[\text{rot}_z\vec{u}_0 \times \vec{u}_0] +$$

$$+ \frac{1}{\sqrt{R}}(\text{rot}_z[\text{rot}_{\xi\eta}\vec{u}_0 \times \vec{u}_0] + \text{rot}_{\xi\eta}[\text{rot}_z\vec{u}_0 \times \vec{u}_0]) +$$

$$+ \frac{1}{R}\text{rot}_{\xi\eta}[\text{rot}_{\xi\eta}\vec{u}_0 \times \vec{u}_0] = \frac{1}{R}\frac{\partial^2}{\partial z^2}(\text{rot}_z\vec{u}_0) + \frac{1}{R^{3/2}}\frac{\partial^2}{\partial z^2}(\text{rot}_{\xi\eta}\vec{u}_0) + \frac{1}{R^{5/2}}\Delta_{\xi\eta}(\text{rot}_{\xi\eta}\vec{u}) \quad (1''')$$

However,

$$\text{rot}_{\xi\eta}[\text{rot}_{\xi\eta}\vec{u}\times\vec{u}] = \text{rot}_{\xi\eta}\left[\text{rot}_{\xi\eta}\begin{pmatrix}0\\0\\\delta\end{pmatrix}\times\begin{pmatrix}0\\0\\\delta\end{pmatrix}\right] =$$

$$= \text{rot}_{\xi\eta}\left[\begin{pmatrix}\frac{\partial\delta}{\partial\eta}\\-\frac{\partial\delta}{\partial\eta}\\0\end{pmatrix}\times\begin{pmatrix}0\\0\\\delta\end{pmatrix}\right] =$$

$$= -\text{rot}_{\xi\eta}\begin{pmatrix}\delta\cdot\frac{\partial\delta}{\partial\xi}\\\delta\cdot\frac{\partial\delta}{\partial\eta}\\0\end{pmatrix} = \begin{pmatrix}0\\0\\\frac{\partial}{\partial\xi}\left(\delta\frac{\partial\delta}{\partial\eta}\right)-\frac{\partial}{\partial\eta}\left(\delta\frac{\partial\delta}{\partial\xi}\right)\end{pmatrix} = \vec{0}$$

Since all the terms in the equation (1'''') that contain $\text{rot}_z$ or $\frac{\partial}{\partial z}$ vanish, we get

$$\frac{1}{R^{5/2}}\frac{\partial}{\partial\tau_1}(\text{rot}_{\xi\eta}\vec{u}_0) = \frac{1}{R^{5/2}}\Delta_{\xi\eta}(\text{rot}_{\xi\eta}\vec{u}_0)$$

Hence

$$\frac{\partial\vec{u}_0}{\partial\tau_1} = \Delta_{\xi\eta}\vec{u}_0$$

or

$$\frac{\partial\delta}{\partial\tau_1} = \Delta_{\xi\eta}\delta(\xi,\eta,\tau)$$

i.e., we got a parabolic equation whose bounded solutions decline exponentially in time $\tau_1 = \dfrac{t}{R^2}$.

Thus, the perturbed solutions of (1") would decline at times $t \sim R^2$.

## 6. Singularity of vorticity at stationary points and strings of singular vorticity

Due to (10) and (12') the quasi-stationary velocity field might be considered as:

$$\vec{u} = C_0(\bar{\Sigma}) \cdot \frac{\mathrm{grad}\,C_0(\bar{\Sigma})}{|\mathrm{grad}\,C_0(\bar{\Sigma})|} + \frac{1}{\sqrt{R}} \vec{\tilde{\delta}}(\bar{\Sigma},\tau) + \frac{1}{\sqrt{R}} \vec{w}_1 \qquad (18)$$

where, as it had been demonstrated in Section 8,

$$\vec{w}_1 = \left|\tilde{\tilde{C}}(\bar{\xi},\bar{\eta},\tau)\right| \cdot \begin{pmatrix} \sin(z + \varphi(\bar{\xi},\bar{\eta},\tau) + \tilde{\varphi}_0(\bar{\xi},\bar{\eta})) \\ \cos(z + \varphi(\bar{\xi},\bar{\eta},\tau) + \tilde{\varphi}_0(\bar{\xi},\bar{\eta})) \\ 0 \end{pmatrix} =$$

$$= \frac{\left|\tilde{\tilde{C}}(\bar{\xi},\bar{\eta},\tau)\right|}{\left|\mathrm{grad}\,C_0(\bar{\xi},\bar{\eta})\right|} \cdot \begin{pmatrix} \dfrac{\partial C_0}{\partial \xi}\cos\tilde{\varphi}_0 + \dfrac{\partial C_0}{\partial \eta}\sin\tilde{\varphi}_0 \\ \dfrac{\partial C_0}{\partial \eta}\cos\tilde{\varphi}_0 - \dfrac{\partial C_0}{\partial \xi}\sin\tilde{\varphi}_0 \\ 0 \end{pmatrix} =$$

$$= \frac{\left|\tilde{\tilde{C}}(\bar{\xi},\bar{\eta},\tau)\right|}{\left|\mathrm{grad}\,C_0(\bar{\xi},\bar{\eta})\right|} (\cos\tilde{\varphi}_0 \cdot \mathrm{grad}\,C_0 + \sin\tilde{\varphi}_0 \cdot \mathrm{ngrad}\,C_0);$$

$$\mathrm{ngrad}\,C_0 = \begin{pmatrix} \dfrac{\partial C_0}{\partial \eta} \\ -\dfrac{\partial C_0}{\partial \xi} \end{pmatrix} \text{ is a normal vector to grad } C_0;$$

$$\overline{\Sigma} = \begin{pmatrix} \overline{\xi} \\ \overline{\eta} \end{pmatrix}$$

$$C_0^2(\overline{\Sigma}) = (A + \gamma_0(\overline{\Sigma}))^2 + \gamma_1^2(\overline{\Sigma})$$

$$\vec{\overline{\delta}}(\overline{\Sigma},\tau) = \begin{pmatrix} 0 \\ 0 \\ \overline{\delta}_1(\overline{\Sigma},\tau) \end{pmatrix}$$

$$\overline{\delta}_1 = -\frac{d\varphi}{d\tau} = -\frac{C_0}{|\mathrm{grad}C_0|}\left(\frac{\partial C_0}{\partial \xi}\cdot\frac{\partial \varphi}{\partial \xi} + \frac{\partial C_0}{\partial \eta}\frac{\partial \varphi}{\partial \eta}\right) - \frac{\partial \varphi}{\partial \tau} =$$

$$= -\frac{C_0 \cdot (\mathrm{grad}C_0, \mathrm{grad}\varphi)}{|\mathrm{grad}C_0|} - \frac{\partial \varphi}{\partial \tau}$$

We are seeking the behaviour of

$$\mathrm{rot}\,\vec{u} = \frac{1}{\sqrt{R}}\mathrm{rot}_{\xi\eta}\left(C_0\cdot\frac{\mathrm{grad}\,C_0}{|\mathrm{grad}\,C_0|}\right) - \frac{1}{R}\mathrm{rot}_{\xi\eta}\vec{\overline{\delta}}_1(\xi,\eta,\tau) + \frac{1}{R}\mathrm{rot}_{\xi\eta}\vec{w}_1$$

in the neighbourhood of a stationary point $(\xi_0,\eta_0)$ of the function $C_0(\xi,\eta)$:

$$\frac{\partial C_0}{\partial \xi}(\xi=\xi_0,\eta=\eta_0) = \frac{\partial C_0}{\partial \eta}(\xi=\xi_0,\eta=\eta_0) = 0$$

Thus

$$C_0(\xi,\eta) = C_0(\xi_0,\eta_0) + \frac{1}{2}(B\tilde{\Sigma},\tilde{\Sigma}) + ... \qquad (19)$$

where

$$\tilde{\Sigma} = \begin{pmatrix} \xi-\xi_0 \\ \eta-\eta_0 \end{pmatrix} = \begin{pmatrix} \tilde{\xi} \\ \tilde{\eta} \end{pmatrix}$$

and $B$ is the Hessian of the function $C_0(\xi,\eta)$ at the point $(\xi_0,\eta_0)$:

$$B = \begin{pmatrix} a & b \\ b & c \end{pmatrix}; a = \frac{\partial^2 C_0}{\partial \xi^2}(\xi_0,\eta_0); b = \frac{\partial^2 C_0}{\partial \xi \partial \eta}(\xi_0,\eta_0), c = \frac{\partial^2 C_0}{\partial \eta^2}(\xi_0,\eta_0),$$ while the truncated

terms are of order $0(\rho^3)$,

$$\rho^2 = \tilde{\xi}^2 + \tilde{\eta}^2,$$

Therefore,

$$\operatorname{grad} C_0(\tilde{\xi},\tilde{\eta}) = B \cdot \tilde{\Sigma} + ... \qquad (20)$$

and

$$\left|\operatorname{grad} C_0(\tilde{\Sigma})\right| = \left| B \cdot \tilde{\Sigma} \right| + ... \qquad (20')$$

Due to the fact that

$$\operatorname{rot}(f \cdot \vec{G}) = f \cdot \operatorname{rot} \vec{G} + [\operatorname{grad} f \times \vec{G}]$$

and

$$\operatorname{grad} \frac{1}{|B\tilde{\Sigma}|} = -\frac{B^2 \tilde{\Sigma}}{|B\tilde{\Sigma}|^3},$$

we (get after several transformations):

$$\frac{1}{\sqrt{R}} \operatorname{rot}_{\xi\eta} C_0 \cdot \frac{\operatorname{grad} C_0}{|\operatorname{grad} C_0|} = -\frac{2C_0}{\sqrt{R}} \cdot \frac{\left[ B^2 \tilde{\Sigma} \times B\tilde{\Sigma} \right]}{|B\tilde{\Sigma}|^3} + ... \qquad (21)$$

Since $B$ is a symmetric matrix we can consider the right-hand side of (21) in the proper basis with the coordinates $(\tilde{\tilde{\xi}}, \tilde{\tilde{\eta}})$:

$$\frac{[B^2\tilde{\Sigma} \times B\tilde{\Sigma}]}{|B\tilde{\Sigma}|^3} = \frac{\lambda_1\lambda_2(\lambda_1 - \lambda_2)\tilde{\tilde{\xi}}\tilde{\tilde{\eta}}}{(\lambda_1^2\tilde{\tilde{\xi}}^2 + \lambda_2^2\tilde{\tilde{\eta}}^2)^{3/2}} \cdot \begin{pmatrix} 0 \\ 0 \\ 1 \end{pmatrix} =$$

$$= \frac{\lambda_1\lambda_2(\lambda_1 - \lambda_2)\sin^2\phi}{\tilde{\rho}(\lambda_1^2\cos^2\phi + \lambda_2^2\sin^2\phi)^{3/2}} \cdot \begin{pmatrix} 0 \\ 0 \\ 1 \end{pmatrix},$$

where

$$\tilde{\rho}^2 = \tilde{\tilde{\xi}}^2 + \tilde{\tilde{\eta}}^2 = \tilde{\xi}^2 + \tilde{\eta}^2 = \tilde{\rho}^2 =$$

$$= (\xi - \xi_0)^2 + (\eta - \eta_0)^2 = \frac{1}{R}((x - x_0)^2 + (y - y_0)^2) =$$

$$= \frac{r^2}{R};$$

$$\tilde{\tilde{\xi}} = \tilde{\rho}\cos\phi$$
$$\tilde{\tilde{\eta}} = \tilde{\rho}\sin\phi \quad .$$

Hence,

$$|(\operatorname{rot}\vec{u})_z| \sim \frac{4C_0(x_0, y_0)}{\sqrt{R}} \cdot \frac{\lambda_1\lambda_2|(\lambda_1 - \lambda_2)\sin^2\phi|}{(\lambda_1^2 + \lambda_2^2 + (\lambda_1^2 - \lambda_2^2)\sin^2\phi)} \cdot \frac{\sqrt{R}}{r} = K(\phi) \cdot C_0 \frac{|\det(B)|^{3/2}}{r} \quad .$$

In the same way, one can prove, that the upward vector

$$\frac{1}{2}\operatorname{rot}_{\xi\eta} \vec{w}_1 = \frac{1}{2}\operatorname{rot}_{\xi\eta} \frac{|\tilde{\tilde{C}}(\overline{\xi}, \overline{\eta}, \tau)|}{|\operatorname{grad} C_0|} (\cos\tilde{\varphi}_0 \cdot \operatorname{grad} C_0 + (\sin\tilde{\varphi}_0) \cdot \operatorname{ngrad} C_0)$$

has singularities of the same type at the stationary points, i.e. the <u>upward component</u> of the vorticity has a singularity of $\dfrac{K}{r}$ type at every stationary point $(\xi_0, \eta_0)$ of the function

$$C_0(\xi,\eta) = C_0\left(\frac{x}{\sqrt{R}}, \frac{y}{\sqrt{R}}\right)$$

where $r = \sqrt{(x-x_0)^2 + (y-y_0)^2}$, $K$ is not depending on the Reynolds number $R$, since we elucidated the zero term in the asymptotic expansion of the large scale vorticity in powers of $\varepsilon = R^{-1/2}$.

In order to assess the $(\xi,\eta)$ - plane component of the vorticity $\frac{1}{R}\operatorname{rot}_{\xi\eta}\vec{\bar{\delta}}_1(\tilde{\xi},\tilde{\eta},\tau)$, one has to investigate the behaviour of $\bar{\delta}_1(\xi,\eta,\tau)$ in the vicinity of the stationary point.

According to (15), $\varphi(\xi,\eta,\tau)$ is a solution of the Cauchy problem :

$$\begin{aligned}\frac{\partial^2\varphi}{\partial\tau^2} &= -\frac{C_0^2(\xi,\eta)}{2}\Delta_{\xi\eta}\varphi - C_0\cdot\left(\frac{\partial C_0}{\partial\xi}\frac{\partial\varphi}{\partial\xi} + \frac{\partial C_0}{\partial\eta}\frac{\partial\varphi}{\partial\eta}\right)\\ \varphi(\xi,\eta,0) &= \operatorname{arctg}\frac{\gamma_1(\xi,\eta,0)}{\gamma_0(\xi,\eta,0)+A}\\ \frac{\partial\varphi}{\partial\tau}(\xi,\eta,0) &= 0\end{aligned} \qquad (15)$$

Here $\gamma_0(\xi,\eta,0) = \gamma_0(\xi,\eta)$ and $\gamma_1(\xi,\eta,0) = \gamma_1(\xi,\eta)$ are small initial perturbations with a finite number of terms in the Fourier expansion.

We can seek the solution of (15) in the vicinity of the stationary point as the series in powers of $\tilde{\xi} = \xi - \xi_0$ and $\tilde{\eta} = \eta - \eta_0$. Then, the zero approximation yields

$$\frac{\partial^2 \varphi_0}{\partial \tau^2} = -\frac{1}{2} C_0(\xi_0, \eta_0) \cdot \Delta_{\tilde{\xi}\tilde{\eta}} \varphi_0(\tilde{\xi}, \tilde{\eta}, \tau)$$

$$\varphi_0(\tilde{\xi}, \tilde{\eta}, 0) = \text{arctg} \frac{\gamma_1(\tilde{\xi}, \tilde{\eta})}{\gamma_0(\tilde{\xi}, \tilde{\eta}) + A} = \quad (15')$$

$$= \text{arctg} \frac{\overline{\varepsilon}\, \tilde{\gamma}_1(\tilde{\xi}, \tilde{\eta})}{\overline{\varepsilon}\, \tilde{\gamma}_0(\tilde{\xi}, \tilde{\eta}) + A} = \overline{\varepsilon}\, \frac{\tilde{\gamma}_1}{A} + \ldots$$

Here $\overline{\varepsilon}$ is a small parameter ("initial noise"). Hence, time dependence is determined by $\gamma_1(\xi, \eta)$ (and thus $\tilde{\gamma}_1(\tilde{\xi}, \tilde{\eta})$). As in Section 6 we consider $\gamma_1(\xi, \eta)$ (and thus $\tilde{\gamma}_1(\tilde{\xi}, \tilde{\eta})$) as a finite double trigonometric polynomial,

$$\tilde{\gamma}_1(\tilde{\xi}, \tilde{\eta}) = \sum_{m,n} \alpha_{mn} e^{i'(m\tilde{\xi} + n\tilde{\eta})},$$

where $m$ and $n$ are bounded. Then

$$\varphi_0(\tilde{\xi}, \tilde{\eta}, \tau) = \frac{\overline{\varepsilon}}{A} \sum_{m,n} e^{i'(m\tilde{\xi} + n\tilde{\eta})} e^{C_0 \sqrt{\frac{m^2 + n^2}{2}} \tau}$$

Thus, the growth of $\varphi_0$ in time is determined by

$$\max_{m,n} \sqrt{m^2 + n^2} = \sqrt{\overline{m}^2 + \overline{n}^2}.$$

Let us consider only the fastest growing term:

$$\varphi_0(\tilde{\xi}, \tilde{\eta}, \tau) = \alpha_{\overline{m}\overline{n}} \frac{\overline{\varepsilon}}{A} \sum_{m,n} e^{C_0 \sqrt{\frac{\overline{m}^2 + \overline{n}^2}{2}} \tau} \cdot e^{i(\overline{m}\tilde{\xi} + \overline{n}\tilde{\eta})} + \ldots$$

Hence

$$\frac{\partial \varphi_0}{\partial \tau} = \sqrt{\overline{m}^2 + \overline{n}^2} \cdot \alpha_{\overline{m}\overline{n}} \frac{\overline{\varepsilon}}{A} \cdot e^{i(\overline{m}\tilde{\xi} + \overline{n}\tilde{\eta})} e^{C_0 \sqrt{\frac{\overline{m}^2 + \overline{n}^2}{2}} \tau}$$

$$\operatorname{grad} \varphi_0 = \sqrt{\overline{m}^2 + \overline{n}^2} \cdot \alpha_{\overline{m}\overline{n}} \frac{\overline{\varepsilon}}{A} \cdot e^{i(\overline{m}\tilde{\xi} + \overline{n}\tilde{\eta})} e^{C_0 \sqrt{\frac{\overline{m}^2 + \overline{n}^2}{2}} \tau} \cdot \begin{pmatrix} \overline{m} \\ \overline{n} \end{pmatrix},$$

i.e.

$$|\operatorname{grad} \varphi_0| = K \cdot e^{C_0 \sqrt{\frac{\overline{m}^2 + \overline{n}^2}{2}} \tau},$$

where $C_0 = C_0(\xi_0, \eta_0)$.

Consider now the $(\xi, \eta)$ - plane component of the vorticity:

$$\frac{1}{R} \operatorname{rot}_{\xi\eta} \overline{\tilde{\delta}}_1 = \frac{1}{R} \begin{pmatrix} \frac{\partial}{\partial \eta} \left( C_0(\tilde{\Sigma}) \cdot \frac{(\operatorname{grad} C_0, \operatorname{grad} \varphi_0)}{|\operatorname{grad} C_0|} \right) + \frac{\partial^2 \varphi_0}{\partial \tau \partial \eta} \\ -\frac{\partial}{\partial \xi} \left( C_0(\tilde{\Sigma}) \cdot \frac{(\operatorname{grad} C_0, \operatorname{grad} \varphi_0)}{|\operatorname{grad} C_0|} \right) - \frac{\partial^2 \varphi_0}{\partial \tau \partial \xi} \end{pmatrix} =$$

$$= \frac{1}{R} \begin{pmatrix} (B\tilde{\Sigma}, \operatorname{grad} \varphi_0) \frac{\partial}{\partial \eta} \frac{C_0(\tilde{\Sigma})}{|B\tilde{\Sigma}|} \\ -(B\tilde{\Sigma}, \operatorname{grad} \varphi_0) \frac{\partial}{\partial \xi} \frac{C_0(\tilde{\Sigma})}{|B\tilde{\Sigma}|} \end{pmatrix} + \ldots$$

where the truncated terms are bounded functions in $\tilde{\xi}$ and $\tilde{\eta}$ for any given $\tau$.

It can be easily checked that

$$\begin{pmatrix} \frac{\partial C_0(\tilde{\Sigma})}{\partial \eta} \\ -\frac{\partial C_0(\tilde{\Sigma})}{\partial \xi} \end{pmatrix} = \tilde{B}\tilde{\Sigma} + \ldots,$$

where $\tilde{B} = \begin{pmatrix} b & c \\ -a & -b \end{pmatrix}$, so that

$$B\tilde{B} = \begin{pmatrix} 0 & ac-b^2 \\ b^2-ac & 0 \end{pmatrix} = \det(B) \cdot \begin{pmatrix} 0 & 1 \\ -1 & 0 \end{pmatrix}$$

Thus, it can be easily seen that

$$\begin{pmatrix} \dfrac{\partial}{\partial \eta} & \dfrac{C_0(\tilde{\Sigma})}{|B\tilde{\Sigma}|} \\ -\dfrac{\partial}{\partial \xi} & \dfrac{C_0(\tilde{\Sigma})}{|B\tilde{\Sigma}|} \end{pmatrix} = -\dfrac{C(\Sigma_0)\det(B) \cdot \begin{pmatrix} \tilde{\eta} \\ -\tilde{\xi} \end{pmatrix}}{|B\tilde{\Sigma}|^3} + \ldots$$

where the truncated terms are bounded for a given $\tau$, and $\Sigma_0 = \begin{pmatrix} \xi_0 \\ \eta_0 \end{pmatrix}$. Let $U$ be the rotation matrix that diagonalizes the symmetric matrix $B$. Hence,

$$\left| \dfrac{1}{R} \mathrm{rot}_{\xi\eta} \vec{\tilde{\delta}}_1 \right| = \dfrac{1}{R} \left| (B\tilde{\Sigma}, \mathrm{grad}\varphi_0) \right| \cdot \left\| \begin{pmatrix} \dfrac{\partial}{\partial \eta} & \dfrac{C_0(\tilde{\Sigma})}{|B\tilde{\Sigma}|} \\ -\dfrac{\partial}{\partial \xi} & \dfrac{C_0(\tilde{\Sigma})}{|B\tilde{\Sigma}|} \end{pmatrix} \right\| =$$

$$= \dfrac{1}{R} \left| (B\tilde{\Sigma}, \mathrm{grad}\varphi_0) \right| \cdot C_0(\Sigma_0) \cdot \dfrac{\det(B)|\tilde{\Sigma}|}{|B\tilde{\Sigma}|^3} = \dfrac{1}{R} \left| (BU\tilde{\Sigma}, \mathrm{grad}\varphi_0) \right| \cdot \dfrac{C_0(\Sigma_0) \cdot \det(B) \cdot |\tilde{\Sigma}|}{|B\tilde{\Sigma}|^3} =$$

$$= \dfrac{C_0(\Sigma_0) \cdot \det(B) \cdot |\tilde{\Sigma}|^2 \left| (BU\vec{e}_1, U\vec{e}_2) \right|}{|B\tilde{\Sigma}|^3} =$$

$$= K \dfrac{e^{c_0 \sqrt{\tfrac{\bar{m}^2+\bar{n}^2}{2}} \tau}}{R} \dfrac{\tilde{\rho}^2 |\lambda_1 \cos\psi + \lambda_2 \sin\psi| \cdot C_0(\Sigma_0) \cdot \det B}{\tilde{\rho}^3 |\lambda_1^2 \cos^2\psi + \lambda_2^2 \sin^2\psi|} =$$

$$\sim \dfrac{K_1(\psi) C_0(\xi_0, \eta_0) \det(B) \cdot e^{c_0 \sqrt{\tfrac{\bar{m}^2+\bar{n}^2}{2}} \tau}}{R \cdot \tilde{\rho}} = \dfrac{e^{c_0 \sqrt{\tfrac{\bar{m}^2+\bar{n}^2}{2}} \tfrac{t}{\sqrt{R}}}}{\sqrt{R}} \dfrac{K_1(\psi) \cdot \det(B)}{r}.$$

where $\vec{e}_1 = \dfrac{B\tilde{\Sigma}}{|B\tilde{\Sigma}|}$ and $\vec{e}_2 = \dfrac{\operatorname{grad}\varphi_0}{|\operatorname{grad}\varphi_0|}$, $\psi$ is the angle between $\vec{e}_1$ and $\vec{e}_2$.

Thus in original ("fast") variables the amplitude of the vorticity plane component in the vicinity of the stationary point $(x_0, y_0)$ is

$$\frac{K_1(\psi) \cdot C_0\left(\dfrac{x_0}{\sqrt{R}}, \dfrac{y_0}{\sqrt{R}}\right) \det(B) e^{c_0 \sqrt{\frac{\overline{m}^2+\overline{n}^2}{2R}}\, t}}{r \cdot \sqrt{R}}$$

where

$$r = \sqrt{(x-x_0)^2 + (y-y_0)^2}\ .$$

Thus, the vertical line $x=x_0, y=y_0$ is a "string" of singular vorticity, while the vorticity lines rotates around it.

## 7. Conclusions: Explicit time dependence and three-dimensionality of the stream as twins emerging from energy density gradient

We consider the Beltrami triplet with variable coefficients (7) as the source for the emergence of large scale streamline tubes for large values of the Reynolds number $R$.

$$u_0 = C_0 e^{-\frac{t}{R}} \cos\varphi(t) \begin{pmatrix} \sin z \\ \cos z \\ 0 \end{pmatrix} + C_0 e^{-\frac{t}{R}} \sin\varphi(t) \begin{pmatrix} \cos z \\ -\sin z \\ 0 \end{pmatrix}$$

$$+ \begin{pmatrix} 0 \\ 0 \\ -\dfrac{\partial\varphi}{\partial t} \end{pmatrix} = \begin{pmatrix} C_0 e^{-\frac{t}{R}} \sin(z+\varphi(t)) \\ -C_0 e^{-\frac{t}{R}} \cos(z+\varphi(t)) \\ -\dfrac{\partial\varphi}{\partial t} \end{pmatrix} \qquad (7)$$

Though (7) clearly indicates the possibility of an upward flow, induced by the presence of the dual Trkal flows with time dependent amplitudes, there is no inherent clue to an equation for the phase. In (7) the absolute value of the velocity in (*x*,*y*)-plane is constant. The breakthrough comes by *variation of* $C_0$ in the (*x*, *y*) –plane: it is now supposed to be a bounded smooth function of *x* and *y* with a small gradient. Such a function has maximal and minimal points. These are the points of the maximum and minimum values of the plane velocity. As it comes out, the liquid flows from the points of the maximal velocity to the points of the minimal velocity, i.e. the liquid flows along the gradient lines of the energy density $C_0$. The distance between these points is determined by the ratio of the velocity change between maximal and minimal values and the average value of the gradient of the function. Since the first number is finite, while the gradient is small, the distance between maximal and minimal points is large. This is how the plane large scale streamlines emerge. The emergence of the unstable upward flow (which is tied to the phase $\varphi(\xi,\eta,\tau)$ between the coupled Trkal flows), i.e. the appearance of the twins-explicit time dependence and three-dimensionality of the flow- is induced by small gradient variation (i.e. by the "initial long wavelength perturbation") of $C_0$ (in fact of $\gamma_1(\xi,\eta,0) = \gamma_1(\xi,\eta)$) in the (*x*, *y*) – plane. It might be seen as the manifestation of the "hydrodynamic instability", which in this case is actually "Eulerian phase instability". The equation for the phase between two dual Trkal flows, which becomes a function of *x* and *y*, might be deduced through a rigorous procedure for the asymptotic expansion for the perturbed solution of the force free Navier- Stokes equation. It comes out that the only possible value of the expansion parameter of the asymptotic expansion consistent with terms of type (14') equals $R^{1/2}$.

Thus, the inverse to the "average" gradient of the function $C_0$, and the distance between maximal and minimal points in the (*x, y*)-plane are close to $R^{1/2}$, as well as the characteristic size of the area in the (*x, y*)-plane. Thus, in case of anisotropic helical solution for the force free Navier-Stokes equation at large Reynolds number **R**, the initial coupling of large scale amplitude modulated dual pair of Trkal (Beltrami) flows together with the orthogonal constant velocity vector form a triplet, which is transformed by a long wavelength perturbation into a large-scale streamline tube, the plane streamlines being stable at times of the order of $R^{1/2}$. These streamlines, which are the gradient lines of the energy density in the orthogonal plane to the anisotropy direction, can be regarded as large scale structures with typical size $R^{1/2}$. The gradient lines connect between "stationary points", where the energy density gradient vanishes. The streamlines inside the domains that do not contain "stationary points" are homotopic. The domains of homotopic plane streamlines are bounded by the so called "separatrices" determining both invariant (under the flow of the liquid) subsets of the plane flow as well as the invariant 3-D polygon prisms ("tubes"); the latter are invariant under velocity and vorticity field flows as well and typically are characterized by the asymptotic collinearity of the velocity and vorticity vectors. The component of the 3-D large scale velocity, that is parallel to the anisotropy direction, is tied to the phase $\varphi(\xi,\eta,\tau)$ between coupled Trkal flows and can be obtained directly as a solution of the Cauchy problem for an elliptic-type equation (the typical case of an ill-posed problem) whose coefficients are determined by the initial conditions. This velocity component outlives the initial Trkal flow and vanishes at times of order $t \sim R^2$.

If we call the initial Trkal flow with a finite amplitude A *the dominant mode*, then the amplitude long wavelength modulation of the dominant mode $\gamma_0(\xi, \eta)$ is responsible for the emergence of the gradient line picture, while the long wavelength amplitude modulation $\gamma_1(\xi, \eta)$ of the dual mode is responsible for the unstable upward flow, i.e. for the emergence of the "twins": explicit time dependence and three-dimensionality of the secondary flow.

Thus, the large scale streamline-vortex tubes are metastable coherent structures. Though the stationary points inside streamline –vortex tubes are singular points of vorticity, the vorticity lines remain inside the tube, while rotating around the "strings"-vertical lines of singularity that are growing from the stationary points in the *x-y* plane.

**References**

1.   Arnold, V., (2007) Topological classification of the trigonometric polynomials related to the affine Coexeter group $A_2$. Proc. Steklov Inst. Math. 258:3-12.
2.   Andreopoulos, Y., (2008), Vorticity and velocity alignment in compressible flows: An experimental study of helicity density in turbulence and vortices, Russian Journal of Electrochemistry, Vol. 44, No. 4, 390-396.
3.   P.R. Baldwin and G.M. Townsend, (1995), Complex Trkalian fields and solutions to Euler's equations for the ideal fluid, Phys. Rev. E 51, 2059- 2068
4.   Yeonatek Choi, Byong-Gu and Chagoon Lee (2009), Alignment of velocity and vorticity and the intermittent distribution of helicity in isotropic turbulence, Phys.

Rev. E80,017309

5. Farge, M., Pellegrino, G. and Schneider, K., (2001), Coherent Vortex Extraction in 3D Turbulent Flows Using Orthogonal Wavelets, Phys. Review Letters,V.87, Number 5, 054501-1-4

6. Levich, E., Tsinober, A., (1983), On the role of helical structures in three-dimensional turbulence, Phys. Lett. A.93, 293-297

7. Levich, E., Tzvetkov, E., (1984), Helical cyclogenesis, Phys. Lett. 100A, 53-56.

8. Levich, E., (2009), Coherence in turbulence: New perspective, Old and New Concepts of Physics, Vol. VI, No.3, 239-457

9. Libin, A., G.I. Sivashinsky and Levich, E., (1987), Long-wave instability of periodic flows at large Reynolds numbers, Phys. Fluids 3, 2984-2986.

10. Libin, A., Sivashinsky, G.I., (1990a), Long wavelength instability of the ABC-flows, Quart. of Appl. Math. Vol. XLVIII, Number 4, 611-623.

11. Libin, A., Sivashinsky, G.I., (1990b), Long wavelength instability of triangular eddies at large Reynolds numbers, Phys. Lett. A 144, 172-178

12. Libin, A., (2010), Large - scale structures as gradient lines: the case of the Trkal flow, Theoretical and Mathematical Physics, Vol.165(2), 1534-1551

13. Lilley, M.S., Lovejoy, S., Strawbridge, K., Schertzer, D., and Radkevich, A., (2008), Scaling turbulent atmospheric stratification, Part II: spatial stratification and intermittency from lidar data, Quart. J. Roy. Meteor. Soc., 134, 301-315

14. Lovejoy, S., Hovde, S., Tuck, A., and Schertzer, D., (2007), Is isotropic turbulence relevant in the atmosphere: Geophys. Res. Lett., 34, L14802, doi:10.1029/2007GL029359.

15. Lovejoy, S., Schertzer, D., Lilley,M., Strawbridge, K., and Radkevich, A., (2008), Scaling turbulent atmospheric stratification, Part I: turbulence and waves, Quart. J. Roy. Meteor. Soc.,134, 277-200.

16. Mininni, P.D., Alexakis, A., Pouquet, A., (2008a), Large scale flow effects, energy transfer, and self-similarity on turbulence, arXiv:physics/0602148v2 [physics.flu-dyn].

17. Mininni, P.D., Alexakis, A., Pouquet, A., (2008b), Non-local interactions in hydrodynamic turbulence at high Reynolds numbers: the slow emergence of scaling laws, arXiv:physics/07809.1939v1 [physics.flu-dyn].

18. Moffatt, H.K., (1985), Magnetostatic equilibria and analogous Euler Flows of arbitrary topological complexity, Part 1 Fundamentals, JFM 159, 359.

19. Moffatt, H.K., (1986), Magnetostatic equilibria and analogous Euler Flows of arbitrary topological complexity, Part 2 Stability Considerations, JFM 166, 359.

20. Mofatt, H.K. (1987), Geophysical and Astrophysical Turbulence in "Advances in Turbulence", Springer Verlag. Berlin, Heidelberg, 1987.

21. Moffatt, H.K., (2001), The topology of scalar fields in 2D and 3D turbulence, Proc. IUTAM Symp. on Geometry and Statistics of Turbulence, Series: Fluid Mechanics and Its Applications, Vol. 59, Kambe, T., Nakano, T., Miyauchi, T. (Eds.), ISBN: 978-0-7923-6711-6.

22. Molinari, J., and Vollaro, D., (2008), Extreme helicity and intense convective towers in Hurricane Bonnie, 136, 4355-4372.

23. Polterovich, L., private communication (2009)

24. Pouquet, Annick, (2011),A few issues in turbulence and how to cope with them, using computers, CU Boulder, July 2011, pouquet@ucar.edu

25. Radkevich, A., Lovejoy, S., Strawbridge, K., Schertzer, D., and Lilley, M., (2008), Scaling turbulent atmospheric stratification, Part III, Space-time stratification of passive scalars from lidar data, Quart. J. Roy. Meteor. Soc.,134, 315-335.

26. Selvam, A.M. (2007), Chaotic Climatic Dynamics, Lunivers Press, Frome, United Kingdom

27. Shtilman, L., and Sivashinsky, G., (1986), Negative viscosity effect in three-dimensional flows, J. de Physique, 47, 1137-1140.

28. Sivashinsky, G., and Yakhot, V.,(1987) Negative viscosity phenomena in three-dimensional flows, (1987) Review A 35(2), 810-820.

29. Sivashinsky, G., (1985), Weak turbulence in periodic flows, Physica D 17, 243-255.

30. Trkal, V., (1918) Časopis št. Mat., 48,302-311

31. Tsinober, A., and Levich, E., (1983), On the helical nature of three-dimensional coherent structures in turbulent flows, Phys. Lett. A 99, 321-324.